\documentclass[5p]{elsarticle}

\usepackage{amsmath}
\usepackage{amssymb}
\usepackage{lineno,subcaption,hyphenat}
\usepackage{diagbox}
\usepackage{url}
\usepackage[dvipsnames]{xcolor}

\usepackage{hyperref}
\hypersetup{pdfstartview={FitH}}

\usepackage{float}
\usepackage{changepage}

\usepackage[most]{tcolorbox}

\newcommand{\boxsection}[1]{%
  \begin{tcolorbox}[
    colback=gray!6,
    colframe=gray!80,
    boxrule=1pt,
    sharp corners
  ]
	\setlength{\parindent}{2em}
  \setlength{\parskip}{0.5em}
    #1
  \end{tcolorbox}%
}

\usepackage{url}
\makeatletter
\g@addto@macro{\UrlBreaks}{\UrlOrds}
\makeatother

\newcommand{\hquad}{\hspace{0.05em}} 

\begin{document}

\begin{frontmatter}

\title{Algorithmic bottlenecks in evolution: Genetic code, symbolic language, and the Great Filter hypothesis}

\author{Mikhail~Prokopenko$^{1,2,3,*}$}
\author{Nihat~Ay$^{3,4}$}
\author{Angelica~Breviario$^{5}$}
\author{Roland~M.~Crocker$^{6}$}
\author{Paul~C.~W.~Davies$^{7}$}
\author{Pauline~Davies$^{7}$}
\author{Darren~Dougan$^{8}$}
\author{Roland Fletcher$^{9}$}
\author{Michael Harr{\'e}$^{1,2}$}
\author{Marcus~G.~Heisler$^{1,10}$}
\author{Zdenka~Kuncic$^{1,11,12}$}
\author{Geraint~F.~Lewis$^{13}$}
\author{Ori Livson$^{1,2}$}
\author{Vivienne Reiner$^{11}$}
\author{Jaime Ruiz Serra$^{1,2}$}

\address{$^{1}$ \hquad The Centre for Complex Systems,  University of Sydney, Sydney, NSW 2006, Australia\\ 
$^{2}$ \hquad School of Computer Science, Faculty of Engineering, University of Sydney, Sydney, NSW 2006, Australia\\
$^{3}$ \hquad Institute for Data Science Foundations, Hamburg University of Technology, Hamburg, 21073, Germany\\
$^{4}$ \hquad Santa Fe Institute, Santa Fe, NM 87501, USA\\
$^{5}$ \hquad Department of Philosophy, Macquarie University, NSW 2113, Australia\\
$^{6}$ \hquad Research School of Astronomy and Astrophysics, Australian National University, Canberra, ACT 2611, Australia\\
$^{7}$ \hquad The Beyond Center for Fundamental Concepts in Science, Arizona State University, Tempe, AZ 85287–0506, USA\\
$^{8}$ \hquad The Big Questions Institute, Surry Hills, NSW 2010, Australia\\
$^{9}$ \hquad School of Humanities, Faculty of Arts and Social Sciences, University of Sydney, Sydney, NSW 2006, Australia\\
$^{10}$ \hquad School of Life and Environmental Sciences, Faculty of Science, University of Sydney, Sydney, NSW 2006, Australia\\
$^{11}$ \hquad School of Physics, Faculty of Science, University of Sydney, Sydney, NSW 2006, Australia\\
$^{12}$ \hquad The Charles Perkins Centre, University of Sydney, Sydney, NSW 2006, Australia\\
$^{13}$ \hquad Sydney Institute for Astronomy, School of Physics, Faculty of Science, The University of Sydney, NSW 2006, Australia\\
$^{*}$ \hquad Corresponding author: mikhail.prokopenko@sydney.edu.au
}

\begin{abstract}
\begin{adjustwidth}{0.5cm}{-9cm}
The Great Filter hypothesis proposes that the emergence of technological societies capable of interstellar travel depends on a small number of exceptionally hard and highly improbable steps. Traditional versions of this hypothesis enumerate such ``hard steps'' along the trajectory from inanimate matter to complex technological societies, but diverge in their explanations for why these particular steps should be so improbable. The theory of Major Evolutionary Transitions also faces challenges in identifying which steps should be considered universally ``hard'' across different evolutionary pathways. In contrast, we argue that two deeply structural obstacles dominate the evolutionary landscape: the coding threshold associated with the origin of the genetic code, and the language threshold associated with the emergence of symbolic communication. We examine the developmental precursors of both transitions and analyze the underlying algorithmic bottlenecks: points at which evolving systems separate code from function, while entangling them within information hierarchies. Using a game-theoretic analysis of coupled signaling and coordination dynamics, we then argue that the corresponding multichannel games may exhibit saddle-type equilibria whose stable manifolds define narrow evolutionary paths, making the transitions intrinsically difficult to traverse. We conjecture that the so-called Great Filter is best understood not as a sequence of isolated improbable events, but as a nested structure of tangled information hierarchies. Under this conjecture, the rarity of advanced societies follows from the difficulty of crossing these coding thresholds in a competitive noisy environment. This hypothesis reframes the Great Filter as an algorithmic property of evolving systems, suggesting that only a small fraction of life may ever traverse the path toward technological societies capable of interstellar travel.
\end{adjustwidth}
\end{abstract}

\end{frontmatter}

\onecolumn

\section{Introduction}

The Great Filter hypothesis is a speculative framework proposed to explain the Fermi Paradox --- the apparent contradiction between the assumed high probability of extraterrestrial life in the Universe and the lack of observable evidence for civilizations\footnote{A more precise term replacing ``civilization'' is ``\emph{a technosignature-bearing society}'' (see Appendix B for this and other italicized terms). An alternative we use henceforth is ``technological society''.} and complex societies~\cite{sagan1963direct,shklovskii1966intelligent,davies2010eerie}.  
The hypothesis suggests that one or more stages in the evolution of life towards societies capable of interstellar travel and long-term large-scale technological activity are extremely unlikely or difficult to pass. This ``filter'' could lie in our past (e.g., the emergence of life, multicellularity) or in our future (e.g., surviving technological self-destruction). If the filter is behind us, humanity is rare and lucky. If it is ahead, our prospects may be grim.

Hanson~\cite{hanson1998great} formalized this idea by identifying nine critical steps in the evolution from inanimate matter to lifeforms capable of interstellar travel and durable astroengineering. According to this view, each step is a potential Great Filter: a stage where progress may stall or fail. These nine steps follow ``an evolutionary path to an explosion which leads to visible colonization of most of the visible universe''~\cite{hanson1998great}:
\begin{enumerate}
	\item The right star system (including planetary conditions): a stable star, habitable zone, and suitable planet.
	\item The emergence of life (abiogenesis): the transition from non-living chemistry to self-replicating life.
	\item Simple (prokaryotic) single-cell life: life that can sustain itself and replicate.
	\item Complex (eukaryotic) single-cell life: cells with internal structures like nuclei and organelles.
	\item Sexual reproduction: the transition which allows for greater genetic diversity and evolutionary adaptability.
	\item Multicellular life: emergence of organisms composed of many specialized cells.
	\item Tool-using animals with intelligence: the transition to organisms which are capable of manipulating the environment and abstract reasoning\footnote{Abstract reasoning is understood as the capacity to construct and manipulate internal models detached from immediate sensory input or context, and reason about hypothetical, non-present, or counterfactual states.}.
	\item Technological society: development of advanced tools, industry, and communication.
	\item Colonization beyond the home planet: expansion into space, detectable across interstellar distances.
\end{enumerate}

This line of reasoning can be traced to the seminal analysis by Carter~\cite{carter1983anthropic}, who argued that the evolutionary emergence of humans depended on successfully traversing a small number of necessary hard (``highly improbable'') steps, including ``the original establishment of the genetic code, and the final breakthrough in cerebral development''.  The core argument of Carter~\cite{carter1983anthropic} is that the existence of human observers sets a limit on how improbable the steps leading to their existence could be, given the available time in the universe and the total lifespan of the sun (our star), which implies that intelligent life might be rare in the universe. In other words, biological evolution is constrained by astrophysical, not just biological, factors.

There are several other proposals describing various ``hard steps'' which were critically examined by a recent study~\cite{mills2025reassessment}. This critical survey distilled five most popular candidate ``hard steps'', coinciding with the exact sequence proposed by Lingam and Loeb~\cite{lingam2019role,lingam2021life}:
\begin{enumerate}[(i)]
	\item The origin of life (``abiogenesis'').
	\item Oxygenic photosynthesis.
	\item Eukaryotic cells (``eukaryogenesis'').
	\item Animal multicellularity.
	\item \emph{Homo sapiens}.
\end{enumerate}
The reassessment of Mills et al.~\cite{mills2025reassessment} led to an alternative model grounded in historical geobiology, in which no hard steps are implied. Instead, \emph{evolutionary singularities}~\cite{de2005singularities} (see Appendix B) were attributed to  mechanisms other than intrinsic improbability, with the timing of human origins governed by the successive opening of global environmental windows of habitability throughout Earth's history~\cite{mills2025reassessment}.

An independent body of work focuses on Major Evolutionary Transitions (METs) --- such as the origin of life,  formation of eukaryotic cells, emergence of multicellularity --- which are characterized by fundamental shifts in how organisms exchange information, cooperate, coordinate, and ultimately survive and replicate~\cite{szathmary1995major,west2015major}. As functional specialization intensifies, these collectives may become tightly integrated, enhancing cooperative efficiency while simultaneously inducing mutual dependence~\cite{west2015major}. Such integration gives rise to higher-order units that increasingly act as the primary targets of selection. By coupling the replication of individual genes to the survival of higher-order structures, selection increasingly favors genes that promote collective viability, thereby enabling greater division of labor and specialization.

Examples of METs include early molecular replicators forming compartmentalized cellular collectives; autono\-mous replicators, likely RNA-based, forming chromosomes to reduce replication-induced information loss; emergence of the genetic code and translation machinery, distinguishing DNA as genetic material and proteins as enzymes; and the emergence of eukaryotic cells with membrane-bound nuclei storing genetic information. This progression continues through multicellularity and eusociality, ultimately extending to language and sociocultural evolution~\cite{szathmary1995major,west2015major,kun2021major,igamberdiev2021drawbridge}. Importantly, these transitions are typically accompanied by the emergence of novel inheritance systems that reshape how information is stored, transmitted, and processed --- sometimes referred to as new ``codes of life''~\cite{barbieri2008mechanisms,kun2021major}. A paradigmatic example is emergence of the genetic code, which introduced a novel inheritance system by mapping nucleotide sequences to amino acids, thereby decoupling information storage from biochemical function and enabling open-ended molecular innovation.

Barbieri~\cite{barbieri2008mechanisms,barbieri2019general} has listed several ``codes of life'', such as signal transduction codes, splicing codes, tubulin code, nuclear signaling code, molecular codes, histone code, transcriptional codes, apoptosis code, bioelectric code, neural codes, etc.~\cite{barbieri2008mechanisms,barbieri2019general}.  However, as pointed by Kun~\cite{kun2021major}, the ``codes of life'' and the major evolutionary transitions represent two distinct components of macroevolution and do not coincide. Two notable exceptions which appear in both lists are \emph{the genetic code} and \emph{human language}. Kun~\cite{kun2021major} argued that for the genetic code and human language the evolution of a ``code of life'' is a MET. 

It has also been argued that major evolutionary transitions occur when systems cross specific ``coding thresholds'', with the saltations opening new evolutionary phase spaces defined by novel information-processing regimes~\cite{woese2004new,kun2021major}. In his analysis of the emergence of DNA, Woese~\cite{woese2004new} argued that
\begin{quote}
``Somewhere along the line there had to have occurred a saltation that we could call the “coding threshold,” where the capacity to represent nucleic acid sequence symbolically in terms of a (colinear) amino acid sequence developed, a development that would generate a truly enormous new, totally unique evolutionary phase space.'' 
\end{quote}

Similarly, a ``language threshold'' may have been crossed when early human groups developed language with symbolic reference and complex syntax\footnote{Here, language denotes spoken symbolic language, while writing, notation and other inscription systems are considered secondary externalizations that amplify and stabilize linguistic control, but do not constitute the language threshold itself. See also subsection~\ref{lelg}.}, which enabled a novel mode of knowledge inheritance and substantially accelerated adaptation~\cite{woese2004new,barbieri2008mechanisms,kun2021major,szathmary2015toward}.  
 As pointed out by Szathm{\'a}ry~\cite{szathmary2015toward}, while animal communication systems mostly include \emph{self-regarding signals} about the immediate and present situation, human language is very different: ``There is a lot of displacement (referring to items that are not present now or are purely imaginary), and it is full of symbolic (arbitrarily conventional rather than indexical or iconic) reference, aided by complex syntax'' (see Appendix B for \emph{indexical} and \emph{iconic references)}. As pointed out by Hauser et al.~\cite{hauser2002faculty}, natural languages extend beyond purely local structure by allowing ``recursive embedding of phrases within phrases'', thereby generating statistical dependencies that can span arbitrarily long distances.
In general, it has been argued that this ``symbolic explosion''~\cite{henshilwood2009reading} coincided with the emergence of syntactic language which evolved within a highly specialized system of social learning~\cite{richerson2000pleistocene} and enabled truly ``semantically unbounded discourse''~\cite{rappaport1999ritual}.

We propose that the Great Filter hypothesis can be interpreted via algorithmic bottlenecks (see next section), offering a perspective to understand constraints on the emergence of intelligent life capable of interstellar travel and astroengineering. Our aim is to explore universal principles that govern specific evolutionary transitions which involve ``coding thresholds''. We hypothesize that these transitions must inevitably pass through algorithmic bottlenecks which may represent saddle-type equilibria in evolutionary dynamics: states that attract trajectories along some directions but amplify perturbations along others, due to environmental, biological, or strategic constraints. We argue that at the bottleneck, the stable manifold of the saddle-type equilibrium occupies only a narrow part of the accessible state space, and trajectories that deviate from it are expelled along unstable directions. As a result, the bottleneck is both difficult to reach and difficult to traverse successfully. In contrast, trajectories that fail to enter these bottlenecks may settle into comparatively stable but stagnant evolutionary regimes, unable to support the emergence of higher-order informational architectures. In other words, if this conjecture is correct, the hardest Great Filter steps may have an algorithmic nature.

\section{Tangled information hierarchies and algorithmic bottlenecks}

How do meaningless molecules create semantically meaningful code? Informally, chemistry in living systems can be identified with hardware, while genetic and epigenetic information can be interpreted as software~\cite{davies1999fifth,walker2013algorithmic}. From this perspective --- \textit{the algorithmic origins of life} --- the emergence of genetic code and accurate translation machinery is crucial, with life having two origins: the hardware and the software. While the spontaneous emergence of chemical complexity proceeds by fairly well understood physical principles (e.g., laws of thermodynamics, energy dissipation, kinetics and mass-action laws, self-assembly due to intermolecular forces), the leap from hardware to software, that is, from local forces and processes to the encoded functional information transmission and global information management, is a much harder step.

\begin{figure}
    \centering
\includegraphics[width=1.0\columnwidth]{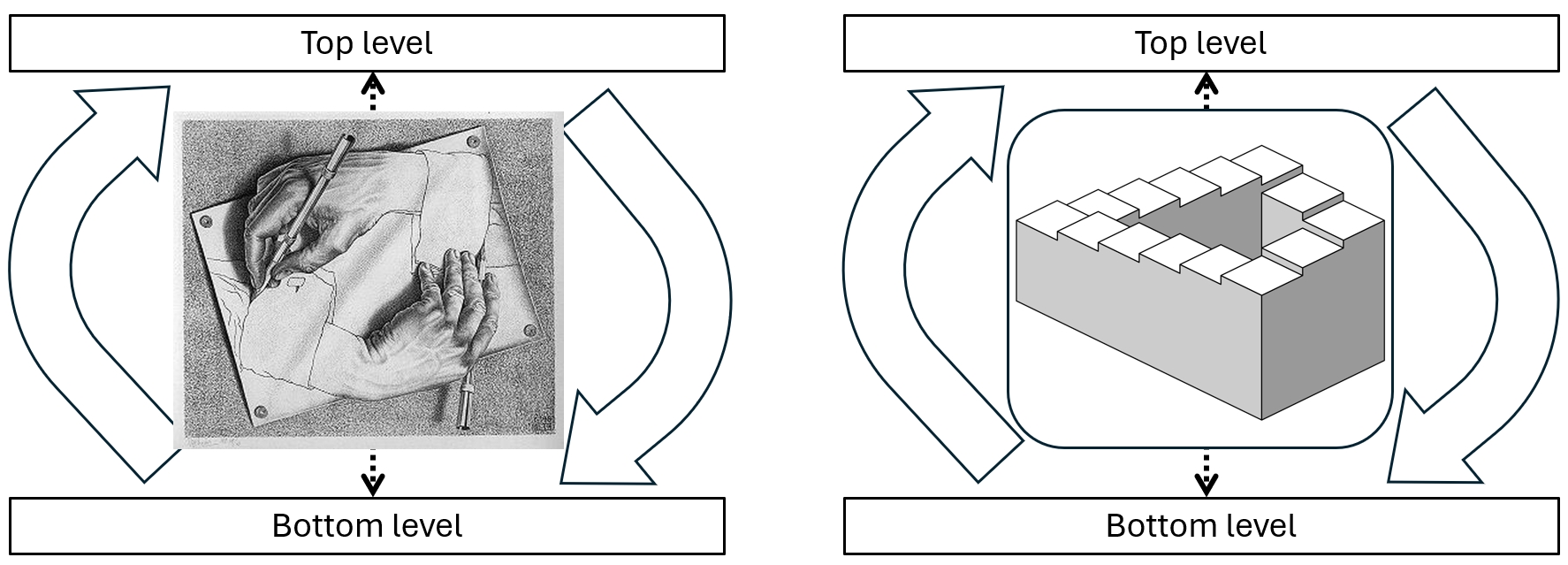}
    \caption{Two well-known strange loops. Left: ``Drawing Hands'' (Escher, 1948)~\cite{wiki-Drawing_Hands}, a visual paradox with two intertwined hands perpetually drawing each other within a recursive loop of drawing an object which draws \textit{itself}. Right: ``Impossible staircase'' (1958)~\cite{penrose1958impossible,wiki-Penrose_stairs}, a spatial pattern connecting the stairs in an impossible way, where a continuous staircase loops back on \textit{itself} within an infinite recursion of ascent and descent.}
    \label{fig:str-loops} \vspace{-3mm}
\end{figure}

Tangled hierarchies are systems in which levels of organization recursively interact, resisting a straightforward hierarchical ordering: ``an interaction between levels in which the top level reaches back down towards the bottom level and influences it, while at the same time being itself determined by the bottom level''~\cite{hofstadter_go_1980}. While a tangled hierarchy describes a structural property of a system, a ``strange loop'' is a dynamic and recursive phenomenon within a tangled hierarchy: ``A strange loop is self-referential, meaning that it loops back on itself in some way. Despite the appearance of movement or progression, a strange loop ultimately returns to its starting point, creating a sense of paradox or contradiction.''~\cite{hofstadter_go_1980}, see Fig.~\ref{fig:str-loops}.

\begin{figure}[t]
    \centering
\includegraphics[width=0.8\columnwidth]{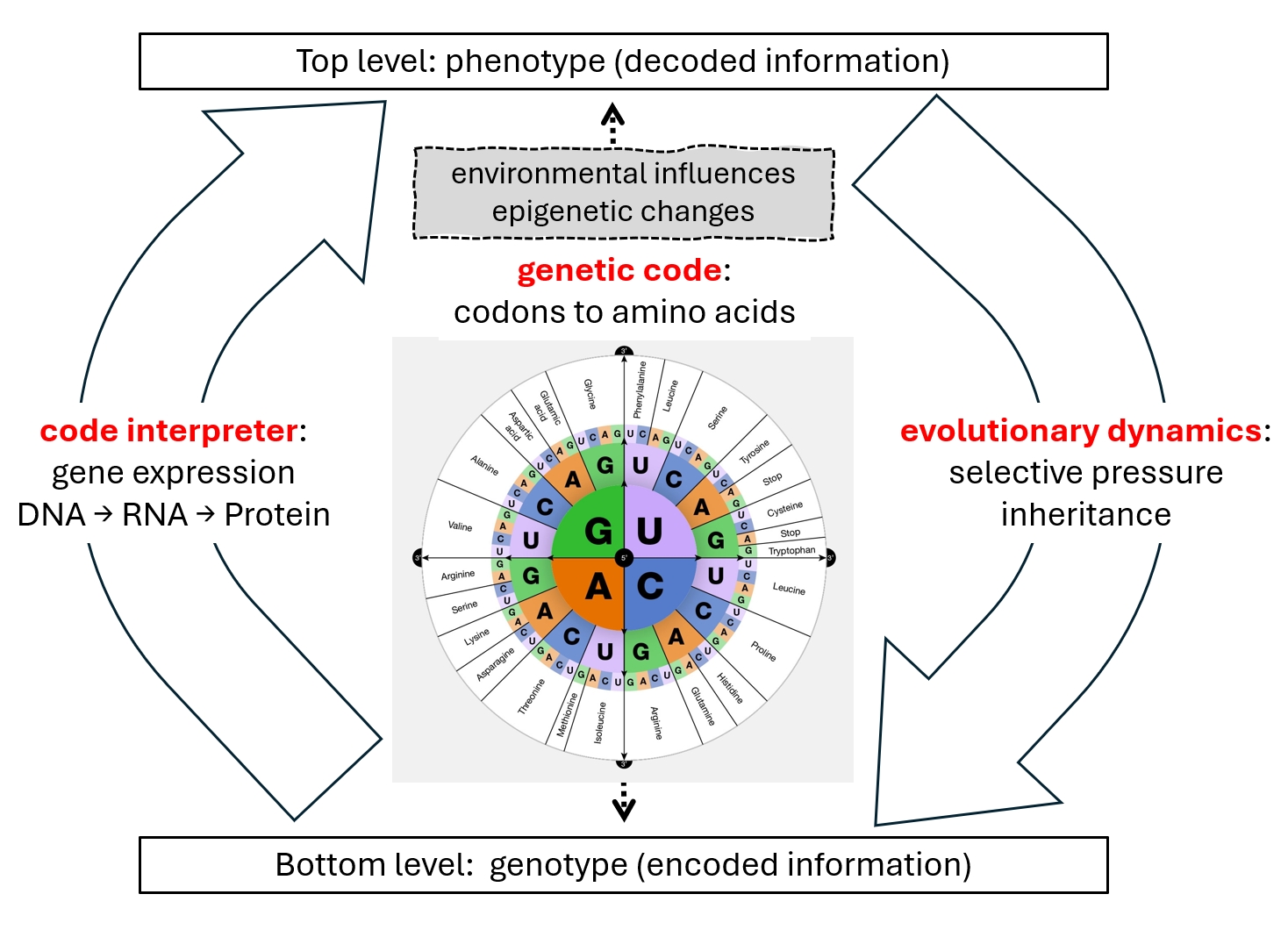}
    \caption{Genotype-phenotype relationship as a tangled hierarchy. Bottom-up construction interprets genetic code via gene expression: transcription and translation of encoded genetic information leads to the decoding of an organism with observable phenotypic traits. Top-down causation: given the phenotype, selective pressure, genetic variation and inheritance influence genotype over generations. Strange loop updates the genetic information which encodes the phenotype \emph{itself}. Genotype-phenotype relationship is realized in context of environmental influences, epigenetic changes and other factors. Genetic code image (public domain): courtesy of the National Human Genome Research Institute (NHGRI)~\cite{NHGRI-gen-code}.}
    \label{fig:gpa} \vspace{-3mm}
\end{figure}

The genetic code encodes the very molecular machinery (tRNAs, aminoacyl-tRNA synthetases, ribosomes) that interprets the code. Given this self-reference, the phenotype-genotype mapping has been interpreted as a ``tangled hierarchy''~\cite{hofstadter_go_1980}, in which phenotypes produced by the bottom-up gene expression influence the genotypes via the top-down selective pressure and inheritance, i.e., via ``downward causation''~\cite{ellis2012top}.  Thus, the evolutionary dynamic constitutes a self-referential loop in which phenotypes determine the evolutionary fate of the genotypes that produced them, see Fig.~\ref{fig:gpa}.

We define an algorithmic bottleneck as a transitional stage in the evolution of an information-managing system where a viable code must be co-established with its interpreter and control machinery, such that (i) the mapping rules, (ii) the interpreting/implementing mechanisms, and (iii) the feedback needed to stabilize their joint operation must all be simultaneously aligned. Because these elements are mutually dependent (the code presupposes an interpreter; the interpreter is specified by the code; stability requires closed-loop control), the system confronts a self-referential and coordination-heavy problem that is hard to traverse and easy to destabilize. Success requires bootstrapping a stable code--interpreter pair under constraints of noise, cost, and ambiguity. 
In this framework, a transition qualifies as an algorithmic bottleneck only when three conditions are jointly satisfied:
\begin{enumerate} 
\item Symbolic encoding: the emergence of an explicit symbolic mapping between representation and function, so that information is partially decoupled from immediate physical function.
\item Interpretive machinery: the emergence of an interpreter capable of executing this mapping, converting encoded information into functional outcomes. 
\item Recursive closure: a self-referential dependency whereby the encoded information contributes to specifying, maintaining, or regulating the interpreter itself.
\end{enumerate} 
Transitions that lack any one of these elements may still be evolutionarily difficult, but they do not constitute algorithmic bottlenecks in the sense intended here. What distinguishes an algorithmic bottleneck is the emergence of a self-referential information-processing architecture that stabilizes an executable mapping between symbolic descriptions and functional outcomes. The core challenge is to reach a coupled fixed point of a joint code--interpreter system ($C, I$). In this system, encoded descriptions $C$ specify an interpreter $I = f(C)$ capable of executing those descriptions by mapping symbolic encodings to organismal dynamics. The interpreter contributes to the construction of an embodied phenotype within which descriptions $C$ are maintained and reproduced. We denote this effective organism-level reproductive dynamics by $g$, yielding $C = g(I)$ and leaving dependence on the phenotype implicit. Here $g$ should not be understood as a direct transformation from interpreter to description. Rather, $g$ is a coarse-grained map representing the organism-level processes, through which descriptions are preserved and reproduced.

Crossing the bottleneck requires the emergence of a recursively closed information architecture, in which $(C, I) = h(C, I)$ for $h(C, I) = (g(I), f(C))$. The bottleneck is crossed when a stable pair $(C^*,I^*)$ emerges such that $(C^*,I^*) = h(C^*,I^*)$. For a coupled fixed-point $(C^*,I^*)$, both code and interpreter are also composite fixed-points $C^* = g(f(C^*))$ and $I^* = f(g(I^*))$ of compositions $g \circ f$ and $f \circ g$, respectively.

Viewed as a search for a coupled fixed point of a self-referential system, traversing an algorithmic bottleneck marks an expansion of the evolutionary search space itself, by introducing a new level of description and interpretation. Given the strong connections between self-reference, fixed-point constructions, and undecidability in formal systems~\cite{yanofsky2003universal,sayama2008construction,hernandez2018undecidability,prokopenko_self-referential_2019}, the emergence of recursively closed code--interpreter systems cannot in general be reduced to a simple local search over predefined functional states~\cite{prokopenko2025biological}. 

The phenotype-genotype mapping provides a canonical example of such a bottleneck. The genetic code establishes a symbolic mapping from codons to amino acids, while the translation apparatus implements that mapping. Yet, the mapping and its interpreter co-specify one another: an encoding cannot function without an interpreter, while the interpreter is itself encoded within the genome (see Appendix A for generic code and interpreter roles). Crossing this threshold requires jointly bootstrapping a stable code--interpreter pair and closing the feedback loops that couple informational rules to biological function. Until the mapping and its implementation become mutually aligned and self-sustaining, the system remains vulnerable to instability and failure. In general, such self-referential, self-modeling dynamics can improve the efficiency of information-processing within a tangled information hierarchy, but they incur the costs of encoding, maintaining and decoding a compressed self-representation~\cite{prokopenko2025biological}.

The self-referential structure of genotype-phenotype organization is closely related to Rosen's concept of closure to efficient causation in relational biology~\cite{rosen1991life}. Rosen argued that living systems are characterized by networks in which the components responsible for functional processes are themselves produced within the same system. This perspective aligns with our characterization of bottlenecks as transitions requiring the co-emergence of a code, its interpreter, and stabilizing feedback loops. Rosen further argued that such organizational closure creates fundamental challenges for purely decompositional representations, suggesting that self-referential biological systems occupy a distinctive position in the space of computable models. Rosen's framework provides a valuable mathematical antecedent for our discussion of self-modeling dynamics and tangled hierarchies. Viewed in this light, the emergence of the genetic code can be understood as the establishment of a causally closed informational architecture in which encoded descriptions and the machinery that interprets them become mutually dependent.

Interestingly, unlike Carter’s original conjecture~\cite{carter1983anthropic}, neither Hanson’s nine-step list~\cite{hanson1998great} nor the five hard steps proposed by Lingam and Loeb~\cite{lingam2019role,lingam2021life} and later evaluated by Mills et al.~\cite{mills2025reassessment} explicitly account for the two unique METs involving ``coding thresholds''. These are the emergence of the genetic code and the emergence of symbolic language.

In discussing transitions in cognitive evolution, Barron et al.~\cite{barron2023transitions} proposed five computational cognitive architectures and related the final transition (a computational architecture capable of ``reflection'') to the emergence of language:
\begin{quote}
``Reflection also opens the door to the efficient use of symbolic languages. ...it is clear that once symbolic languages are in play new forms of reflection then become possible, offering further efficiencies in task acquisition and execution. Language can be seen as a way to compress instructions into simpler forms [89] and enable new types of mental programmes [9]. By a ‘programme’, we mean any way in which control flow can be effectively represented and manipulated, thereby shifting the burden of developing new control flow from the biological to cognitive.''
\end{quote}
This characterization suggests why the language threshold may constitute an algorithmic bottleneck: symbolic language enables compressed functional information to serve as executable control flow, creating a self-modeling loop that entangles cognitive ``software'' with its neural ``hardware''. 

\begin{figure}[t]
    \centering
\includegraphics[width=0.8\columnwidth]{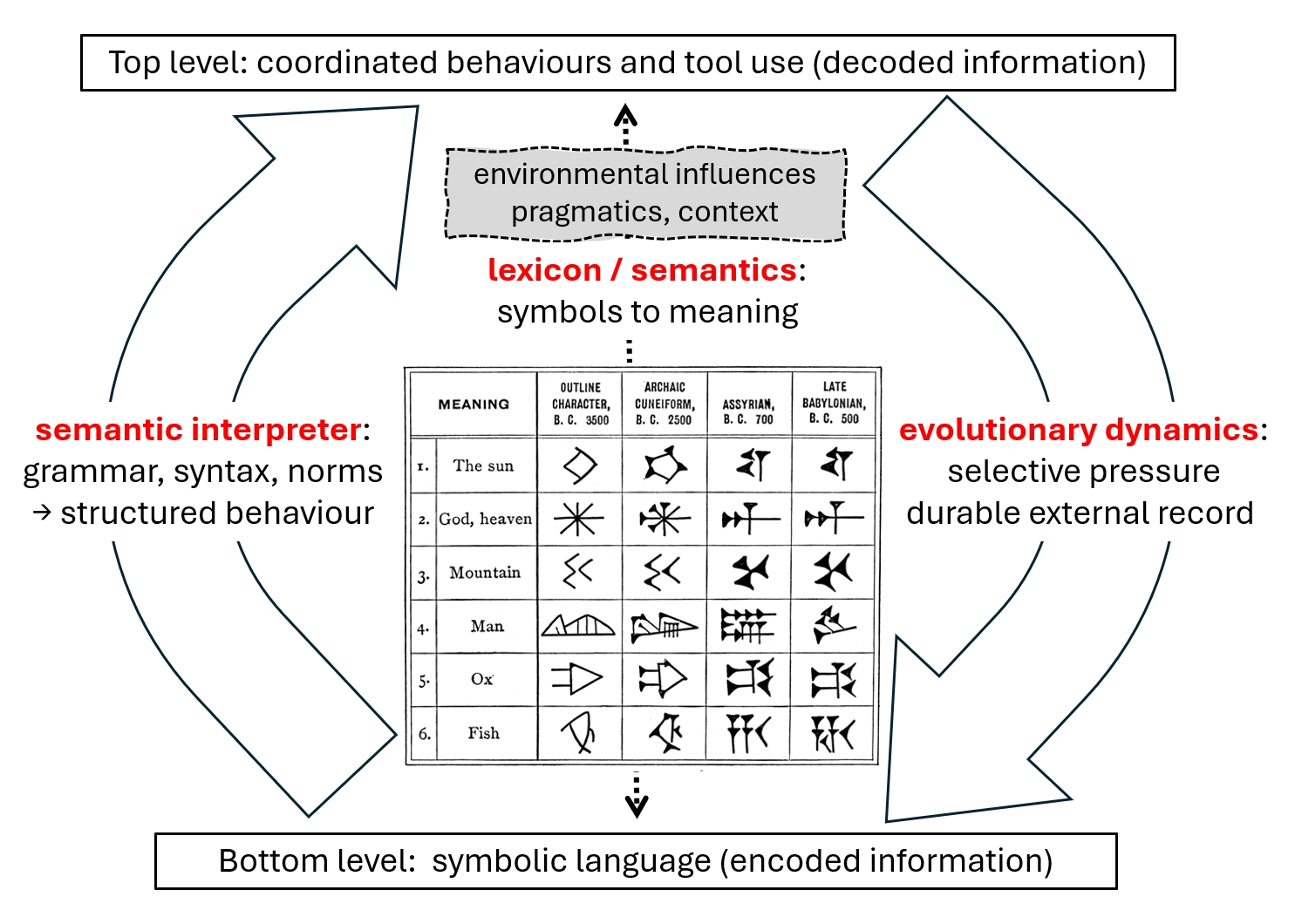}
    \caption{Language-behavior relationship as a tangled hierarchy. Bottom-up construction interprets symbolic language via grammatical and semantic machinery: grammar, recursive syntax, conversational norms and conceptual schemes shape the decoding of communicative acts, coordinated behaviors and tool use. Top-down causation: given the social behavior, selective pressure, linguistic innovation and external storage influence language over generations. Strange loop updates the symbolic information which encodes the social behavior \emph{itself}. Language-behavior relationship is realized in context of environmental influences, pragmatics and contextual interpretation, among other factors. Cuneiform image (copyright-free): a table illustrating evolution of cuneiform~\cite[p. 55]{myers1904ancient}.}
    \label{fig:sla} \vspace{-3mm}
\end{figure}

Passing the language threshold marks the stage at which symbolic control flow ceases to be confined to individual cognition and becomes distributed across external artifacts and social practices\footnote{This characterization does not assume the extended mind thesis~\cite{clark1998extended}, remaining agnostic about whether such external structures constitute cognition itself.}. This shift allows recursive control structures to stabilize and compound across time, enabling cumulative, trans-generational computation (see Fig.~\ref{fig:sla}). While other species, including bees or dolphins, may exhibit sophisticated, context-dependent communication, human language uniquely supports the external storage and recursive manipulation of symbolic structures outside of individual brains. 
The relevant distinction is not a conjunction of species-specific adaptations, but the coupling of two abstract informational functions: durable external symbolic memory and recursive symbol manipulation. These functions are substrate-independent and, in principle, would be equally significant whether realized by humans, dolphins, or extraterrestrial intelligences.

Importantly, both transitions --- the emergence of the genetic code and the emergence of symbolic language --- go beyond emergent cooperation, division of labor and collective coordination: they require the establishment of an explicit mapping, or ``code''. This creates an algorithmic self-modeling problem shared by both transitions:
\begin{enumerate}[I.]
	\item The emergence of genetic code and translation machinery: progression from an early (e.g., RNA or protein/peptide) world to a DNA/protein world, in which heritable information specifying phenotypic function is stably encoded in DNA and interpreted (decoded) via a dedicated translation apparatus. 
	\item The emergence of symbolic language and linguistic interpretive machinery: progression from pre-linguistic communication and behavioural cues to a symbol-based representational system, in which abstract meanings are stably encoded in symbols and interpreted (decoded) via shared semantic and syntactic rules.
\end{enumerate}
Both these transitions mark a progression from locally constrained, context-bound signaling to  increasingly explicit symbolic codes supported by collective mechanisms for parsing, interpretation, and meaning reconstruction. 
Just as the genetic code required co-evolution of the translation machinery that maps nucleotide sequences into functional proteins, symbolic language needed the joint stabilization of symbolic reference and syntax.  Importantly, the relative purity of the genetic code, its almost context-independent mapping between codons and amino acids, contrasts with the inherently context-sensitive nature of human language. Yet what is shared across both transitions is not full context-independence, but the emergence of conventionalized symbol systems whose interpretation is stabilized by communal processes, allowing meaning to persist across contexts even when the mechanisms differ. Appendix A elaborates on the analogy between genetic and linguistic coding systems in more detail.

Our use of tangled hierarchies is related to formal approaches such as Causal Emergence Theory (CET), which quantifies the causal efficacy of macroscopic descriptions using information-theoretic measures tracking interventions and predictability~\cite{hoel2017map,hoel2026quantifying}. Similarly, studies of collective computation and coarse-graining in adaptive social systems have shown how emergent macro-level variables can both summarize and regulate lower-level interactions through feedback loops~\cite{flack2017coarse}. Other studies quantified a form of circular causality where neural activity and extracellular electric fields continuously shape each other~\cite{a2026ephaptic}. These approaches provide a rigorous account of how higher-level organizational states may exert top-down constraints on lower-level dynamics. Importantly, we do not claim that these frameworks directly explain the emergence of the genetic code or symbolic language. Rather, they provide formal and computational precedents for our broader claim that higher-order informational organizations can acquire causal efficacy and participate in stabilizing evolutionary dynamics.

\section{Communication and Cooperation}

A central premise of the ``algorithmic origins of life'' framework for explaining the emergence of the genetic code~\cite{walker2013algorithmic} is the onset of global information management, whereby encoded functional information --- ``software'' --- acquires causal efficacy over the underlying physical substrate, or ``hardware''. In this section, we examine key aspects  of this self-referential dynamic in the context of (a) the emergence of genetic code, ``the lingua franca of genetic commerce''~\cite{woese2000interpreting,woese2004new}, and (b) the emergence of symbolic language with its ``unlimited hereditary potential''~\cite{szathmary2015toward}.

Because both genetic coding and symbolic language formation require agents (molecules, cells, or individuals) to coordinate using shared conventions, we will adopt \emph{evolutionary game theory} to analyze relevant signaling games (sender–receiver alignment) and coordination games (alignment of resource use). This approach will clarify why misalignment, ambiguity, and free-riding (defection) can be evolutionarily attractive locally yet destructive globally. In turn this will expose the fundamental tension that makes an algorithmic bottleneck particularly difficult to traverse. We will argue that this is not a problem of isolated optimization, but rather a strategic coordination problem involving multiple agents dynamically engaged in both competitive and cooperative interactions across multiple channels.

A canonical two-player game with two strategies, Cooperate (C) and Defect (D), can be modeled with a payoff matrix for two players (Row and Column), using  four payoff parameters (see Table~\ref{tab2} and Box 1):
\begin{itemize}
\item $R$ (Reward): payoff to each if both cooperate
\item $T$ (Temptation): payoff to a defector against a cooperator
\item $S$ (Sucker's payoff): payoff to a cooperator against a defector
\item $P$ (Punishment): payoff to each if both defect
\end{itemize}

\begin{table}[t]
\centering
\caption{Payoff matrix for two players (Row and Column), using four payoff parameters.}
\begin{tabular}{|l|c|c|}\hline
\diagbox[width=2.7cm, height=0.8cm]{Row}{Column} & {Cooperate} & {Defect} \\
\hline
Cooperate & \diagbox[width=2.7cm, height=0.8cm]{$R$}{$R$} & \diagbox[width=2.7cm, height=0.8cm]{$S$}{$T$} \\
\hline 
Defect    & \diagbox[width=2.7cm, height=0.8cm]{$T$}{$S$} & \diagbox[width=2.7cm, height=0.8cm]{$P$}{$P$} \\
\hline
\end{tabular}
\label{tab2}
\end{table}


\

\


{\textbf{Box 1: Canonical game-theoretical models.}}

\boxsection{

Many standard games are obtained by imposing different orderings on $T$, $R$, $P$, $S$, e.g., Prisoner's Dilemma or Stag Hunt game (see below). Nash equilibrium is a key concept where each player chooses a strategy that is their best response to the strategies chosen by other players, resulting in a game-theoretically stable outcome where no single player can improve their payoff by changing their strategy alone, assuming others keep theirs constant. In Nash equilibrium no player could gain more by changing their decision given what others did, although it does not always lead to the best overall result for the group. 

In Prisoner's dilemma, each player simultaneously selects an action without communication or binding commitments. Using the canonical payoff parameters, the model assumes that $T > R > P > S$. Mutual cooperation yields a higher payoff for both than mutual defection, yet unilateral defection strictly dominates cooperation for each player. As a result, the Prisoner's Dilemma has a single pure-strategy Nash equilibrium where both players defect, despite mutual cooperation being \emph{Pareto optimal}. 
This tension between individual rationality and collective optimality makes the Prisoner’s dilemma a foundational example in the analysis of strategic interaction, equilibrium selection, and social dilemmas~\cite{axelrod1981evolution}.

Stag Hunt game is a canonical game-theoretic model with two or more agents which can choose between strategies of hunting a Hare (i.e., acting alone which is safe and has minimal risk, but offers a low payoff), or hunting a Stag (i.e., requiring cooperation with others which is risky because the hunt fails if others defect, but offers high payoff). Using the canonical payoff parameters, the model assumes that $R > T \ge P > S$. 
The Stag Hunt game differs from the Prisoner's Dilemma in its strategic structure: it admits two pure-strategy Nash equilibria: mutual cooperation and mutual defection. This distinction highlights the coordination challenge in the Stag Hunt, where trust and assurance are required to reach the cooperative equilibrium, whereas the Prisoner's Dilemma is dominated by incentives to defect regardless of the other player's choice. Importantly, the mutual defection is the Stag Hunt game is \emph{risk-dominant} being less risky, while mutual cooperation is suffering from uncertainty due to a lack of trust and the fear of the worst-case scenario. 

Evolutionary game theory applies game theory to evolving populations, modeling how strategies (behaviors, traits) spread or die out based on their fitness payoffs, which depend on the strategies of others in the environment (the population itself). Unlike classical game theory, evolutionary game theory focuses on frequency-dependent selection, explaining adaptation and the evolution of cooperation, altruism, and complex behaviors in biology, economics, and sociology. A refinement of the Nash equilibrium that remains stable under evolutionary dynamics is evolutionarily stable strategy (or set of strategies): a behavioral strategy that, once adopted by a population, cannot be successfully invaded or displaced by any rare, alternative ``mutant'' strategy (or set of strategies), ensuring its stability over time.
}


\subsection{Emergence of genetic code}
\label{egc}

The evolutionary problem underlying the emergence of genetic code has been succinctly formulated by Woese~\cite{woese2004new}: ``how the genotype-phenotype relationship had come to be''.  This question leads us to consider the precursors of the genetic coding threshold, i.e., the transition to symbolic nucleic-acid--protein coding, focusing on how early molecular interactions and collective dynamics could overcome the algorithmic bottleneck inherent in establishing a self-consistent genotype–phenotype mapping.

\subsubsection{Precursors of the genetic ``coding threshold''}

Before a stable genotype-phenotype map could crystallize, several enabling conditions had to co-emerge. On the one hand, informational polymers or other heritable chemical regularities needed sufficient templating fidelity and length. This supported heritable variation and sustained polymerization and compartmentalization, helping to concentrate reactants and protect nascent networks~\cite{eigen1971selforganization,szostak2001synthesizing}. On the other hand, catalytic specificity had to rise above background enzyme promiscuity, enabling mechanisms that could associate informational patterns with functional outcomes~\cite{khersonsky2006enzyme,rodin2008origin}, and thus, introduce the rudiments of a code–interpreter pair. Additionally, control and feedback processes such as proofreading, repair, and kinetic discrimination, were needed to suppress error cascades~\cite{hopfield1974kinetic,eigen1989molecular}. Finally, population-level processes, such as horizontal exchange, communal selection, and convention formation, were required to align partially compatible mappings and drive convergence toward shared assignments under pressures for error minimization, redundancy, and compositionality~\cite{woese1965evolution,vetsigian2006collective}. Together, these elements established the pre-coding scaffold: a milieu in which an initially rudimentary and easily lost mapping between sequence and function could bootstrap into a stable, scalable code--interpreter system. These shared functional prerequisites can be approached from different explanatory perspectives (see Box 2).

\

\


{\textbf{Box 2: Information and metabolism in early life.}}

\boxsection{

One resolution to the genotype-phenotype duality is that the modern DNA–protein system evolved from a simpler precursor in which a single molecular species performed both informational and catalytic roles. In contemporary biology, RNA mediates between DNA and proteins, and is unique in its ability to function as both a genetic polymer and an enzyme. These properties underpin the influential RNA-world hypothesis, which proposes that all known life descended from an ancestral population that relied on RNA as its sole major biopolymer before the emergence of DNA and proteins~\cite{gilbert1986origin,joyce2002antiquity,gianni2026small}. 

The RNA-world hypothesis holds that the genetic code emerged within an RNA-based system in which RNA molecules acted as both informational templates and catalytic ribozymes capable of discriminating between amino acids.  RNA replication provided the substrate for encoding rules, while proto-tRNAs and proto-ribozymes mediated primitive codon-amino acid associations. Coding then evolved through refinement of translation, driven by selection for greater accuracy, enzymatic diversity, and replicative fitness.

Thus, it is assumed that during the early era of nucleic-acid–based life, primitive cell-like entities may have existed as loosely connected conglomerates open to extensive horizontal gene transfer, with evolutionary dynamics characterized by ``communal descent''~\cite{woese1972emergence,vetsigian2006collective}. Crucially, at the coding threshold the genetic code functioned ``not only as a protocol for encoding amino acid sequences in the genome but also as an innovation-sharing protocol''~\cite{vetsigian2006collective}: it served both as a component of the replication machinery and as a medium for encoding and disseminating information about the environment. Different proto-cells may have discovered distinct adaptive innovations, and the horizontal exchange of such innovations was facilitated by a shared, converging proto-code.

In contrast, metabolism-first models place chemical reaction networks, rather than genetic molecules, at the origin of biochemical organization. The genetic code appears only after metabolic cycles become stable and catalysts diversify, with the selection favoring more reliable inheritance mechanisms. According to this view, autocatalytic networks function without symbolic representation, and simple peptides, minerals, or small molecules drive early metabolism which precedes informational templates~\cite{dyson1985origins,wachtershauser1988before,kauffman1993origins}. The genetic code then arises to orchestrate and stabilize pre-existing metabolic chemistry, rather than to enable early replication. Thus, the mapping from codons to amino acids is shaped not by ribozyme discrimination but instead, by biosynthetic pathways, chemical affinities and metabolic availability of amino acids. 
}


\newpage

A principal disagreement between these perspectives is that coding in the RNA-World is an intrinsic requirement of an RNA-centric replicative system, while in the metabolism-first scenario, coding is an extrinsic solution, emerged to stabilize and propagate something more fundamental: metabolism. In comparing these two approaches, Walker and Davies~\cite{walker2013algorithmic} argued that neither the genetics-first (``digital'') nor metabolism-first (``analogue'') trajectories provide a satisfactory
explanation, highlighting that while information may be expressed in specific chemical structure (digital or analogue), it lacks autonomy~\cite{walker2013algorithmic}. Instead, their study shifted attention toward the underlying logical architecture of information processing and control, as well as the pathways of information flow in living systems --- pathways that are themselves influenced by information encoded within the system. Under the algorithmic interpretation of Walker and Davies~\cite{walker2013algorithmic}, the threshold is crossed when ``an abstract and non-physical systemic entity (algorithmic information) effectively becomes a causal agent capable of manipulating its material substrate``, giving rise to a form of distributed information control~\cite{walker2013algorithmic}.

\subsubsection{The codon-signaling game}

It is instructive to consider plausible scenarios for the emergence of genetic code in a game-theoretic context (see Appendix B for \emph{canonical} and \emph{evolutionary games}). In general, signaling game theory examines strategic interactions where a sender communicates information to a receiver through signals that may be truthful, partial, or misleading~\cite{zahavi1975mate}. The existence of information asymmetry among the macromolecules, including molecular deception, was studied using sender–receiver signaling games, providing evidence that selfish behavior differentiates biochemistry from abiotic chemistry~\cite{massey2018origin}.  Massey and Mishra \cite{massey2018origin} identified the genetic code as a signaling convention ``between genes (senders), their mRNAs (the signal) and the set of tRNAs in the cell (receiver)'', while pathogens and selfish elements were argued to subvert this convention~\cite{massey2018origin}.

In a pre-DNA/protein environment, early molecular systems may be interpreted as players competing for replication and catalytic efficiency, with various co-evolving coding schemes and molecular interactions. 
In the evolutionary codon-signaling game, players are ``senders'' (encoding information, e.g., mapping RNA codons to amino acids) and ``receivers'' (decoding information, interpreting codons and acting: tRNA or ribosome assigning amino acid)~\cite{prokopenko_stigmergic_2009,obst2011origins}. 

Agents (e.g., molecular replicators such as proto-ribo\-zymes, RNA strands, peptides) interact under payoff structures which favor reliable information transfer and functional catalysis. For example, payoff can be formed as the effective protein yield, adjusted for penalties due to amino-acid \emph{misincorporation} (i.e., translational decoding errors) and the translation time costs. Then, high payoff may be achieved when sender and receiver coordinate on a stable, functional mapping, while low or negative payoff is attained under mismatches or noisy, slow interpretations. Game strategies may involve random codon assignments, autocatalytic loops, template-based replication, and so on. 

The number of possible codon-to-amino acid mappings is, however, combinatorially vast, and the agents may fail to align on a shared coding scheme, leading to fragmented or incompatible coding systems. High mutation rates and translational errors further disrupt reliable genotype-pheno\-type mapping: mutations perturb the sender's message (codon sequence), while mistranslations undermine the receiver’s interpretation (decoding). Furthermore, limited availability of nucleotides, amino acids, or catalytic surfaces may favor short-term survival strategies over long-term code optimization. 

The selective pressure for refining the code is further reduced by weak or delayed feedback between coding accuracy and functional performance, allowing suboptimal mappings to persist. Since the mapping from code configurations to fitness is highly non-linear and discontinuous (i.e., the game has a rugged fitness landscape), it is difficult for evolutionary dynamics to traverse the landscape valleys toward globally optimal codes. In other words, early arbitrary assignments (i.e., local optima) may be entrenched, preventing exploration of potentially superior coding schemes.

\begin{figure}
    \centering
\includegraphics[width=1.0\columnwidth]{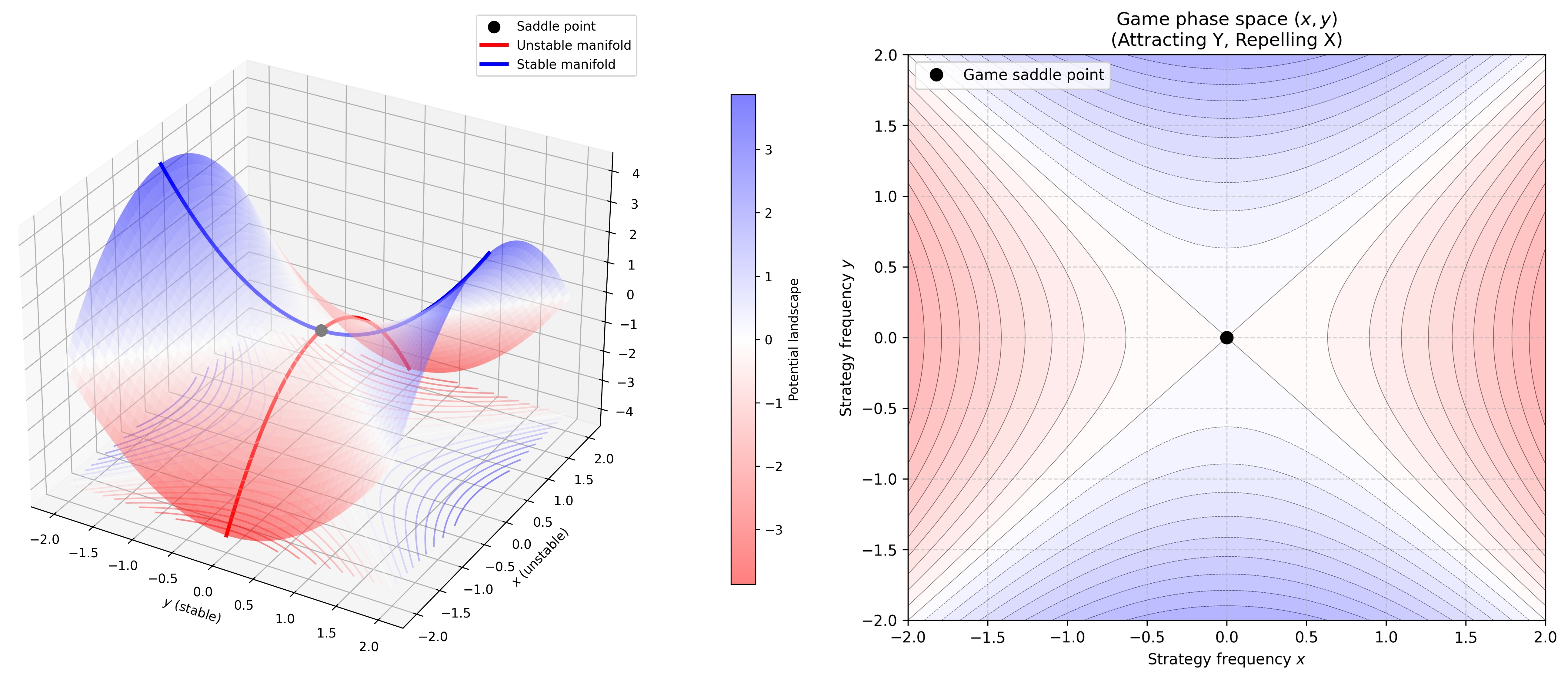}
    \caption{Left: A generic two dimensional saddle-point manifold at the intersection of the red and blue curves. Along the blue curve the saddle point is a local minimum, and along the red curve it is a local maximum. Right: In game dynamics, trajectories converge along stable manifolds (local attraction) but face rapid repulsion along unstable manifolds (repulsion paths), meaning that small strategy deviations take the system away from the equilibrium.}
    \label{fig:saddle}
\end{figure}

In this scenario,  the dynamics may be organized around a saddle-point manifold rather than an attractor. Evolutionary pressures can guide populations along its stable manifold toward a narrow bottleneck of coding coherence, but the corresponding unstable manifold acts as an exit channel through which translation errors, mutational drift, or non-cooperative strategies can rapidly amplify perturbations and drive the system away from viability. Figure~\ref{fig:saddle} illustrates a saddle-point manifold as a game-theoretic equilibrium. Such dynamics make the apparent convergence deceptive: the same evolutionary forces that funnel populations toward coherence can leave them poised near an instability from which error catastrophes, translational noise, or parasitic strategies readily trigger collapse.

\subsubsection{The metabolic cooperation game}

The evolutionary dynamics of the codon-signaling game considered in the previous subsection makes convergence toward a widely shared code unlikely without strong constraints or coordination mechanisms, for example, provided by metabolism. In early Earth environments, molecular replicators needed to develop metabolic pathways to harness energy and replicate. These pathways often involved shared enzymes or metabolites that diffused into the environment, benefiting nearby cells~\cite{prokopenko_stigmergic_2009}.

Game-theoretically, metabolic cooperation can be interpreted as a coordination game: a simultaneous game in which players (e.g., proto-cells or cells) benefit from selecting the same strategy, as mutual alignment yields higher payoffs than mismatched choices. Specifically, one may consider a spatial public goods game with diffusion constraints (see Box 3). 

\

\


{\textbf{Box 3: Public goods game.}}

\boxsection{

Public goods game models social dilemmas where individuals choose between contributing to a shared resource or free-riding for personal gain.  In the basic game, participants simultaneously and secretly decide how many of their endowed tokens to contribute to a public pot. Each player’s payoff consists of private consumption --- defined as their remaining endowment after contribution --- plus returns from the public good, equal to the total contributions scaled by a fixed multiplier. The game demonstrates the conflict between self-interest and collective benefit, often resulting in under-provision of the good, unless mechanisms like punishment or rewards are introduced. 

In a spatial public goods game, agents are arranged on a spatial network (e.g., a lattice) rather than form a fully mixed group, and their ability to contribute and receive benefits is constrained by the presence and actions of their immediate neighbors. This spatial structure may introduce diffusion, so that contributions and strategies spread or are influenced by local interactions, rather than the global, uniform conditions of a classic public goods game~\cite{szolnoki2012conditional}. Diffusion limits the reach of public goods, making local cooperation more viable.

In an evolutionary public goods game coupled with ecological dynamics, outcomes may be highly sensitive to initial conditions (shaping a saddle–point structure of the dynamics). Dependent on the parametrization, this may lead to oscillations or population extinction~\cite{hauert2006evolutionary}.
} 


\

\

While cooperators produce and share essential metabolites or enzymes, defectors consume shared resources without contributing. The agents interact under different payoffs: cooperators incur a cost but increase group fitness; defectors gain short-term advantage but risk system collapse. In a \emph{public goods game} with $n$ agents, there are at least $k \le n$ contributing cooperators which are necessary to produce the public good. The cooperation dilemma arises because each individual has an incentive to avoid the cost of contributing resources (``volunteering'') and instead free-ride on the resources contributed by others (``public goods''). Yet, if the public good is not provided, all individuals incur a cost greater than that of volunteering~\cite{archetti2012game}. However, since $n$-person public goods games aggregate contributions non-linearly, a cooperation is possible, and a polymorphic state in which cooperators and non-cooperators coexist can be stable~\cite{archetti2012game}.  

In selection-diffusion models of ecological public goods, cooperation emerges from spatial pattern formation. Slow diffusion causes cooperators to cluster in productive patches (activation: the presence of nearby cooperators enables or amplifies cooperative payoffs). Rapid diffusion enables defectors to invade and exploit these patches (inhibition: the presence of nearby defectors, or an insufficient density of cooperators, suppresses or undermines cooperative payoffs). The tension between these processes yields a variety of patterns resembling Turing patterns, including persistent coexistence and spatial chaos~\cite{wakano2009spatial}. In these models, stability hinges on patch formation and invasion fronts (activation-inhibition), effectively turning point equilibria into geometric conditions on cooperative clusters, with the pattern formation shaped by the antagonistic forces of cooperators (activators) and defectors (inhibitors)~\cite{wakano2009spatial}. 

From a dynamical-systems viewpoint, the interior coexistence equilibrium of cooperators and defectors may also be a saddle-type equilibrium with stable and unstable manifolds, such that some perturbations decay whereas others grow~\cite{hauert2006evolutionary,hauert2008ecological}. In the spatial selection-diffusion extension, the corresponding homogeneous coexistence state can become unstable to spatial perturbations, leading to diffusion-driven pattern formation and acting as an organizing center for pattern-forming transitions~\cite{wakano2009spatial}. Thus, without spatial clustering or regulatory mechanisms (e.g., quorum sensing, membrane compartmentalization), cooperation cannot stabilize.

\subsubsection{Algorithmic bottleneck in the emergence of the shared genetic code}

Crucially, the spatial public goods game and the codon-signaling game can be viewed as coupled through a \emph{multichannel} feedback loop, in which the output of each serves as an input to the other. For instance, the outcome of the signaling game --- a more robust genetic code --- can regulate quorum sensing and stabilize cooperative behavior, thereby increasing the collective payoff in the public goods game. In turn, the public goods game can reinforce the universality of the genetic code within spatial clusters by selectively rewarding compatible signaling systems. If, however, the codon-signaling game yields only a partially coherent genetic code, permeability of the protocell's membrane  may be impaired, reducing metabolite retention, and genetic regulation may fail to suppress freeloading. As a result, a spatial public goods game with a saddle-type polymorphic equilibrium may shift toward defector dominance, leading to the collapse of the shared metabolic infrastructure.

\begin{figure}
    \centering
\includegraphics[width=1.0\columnwidth]{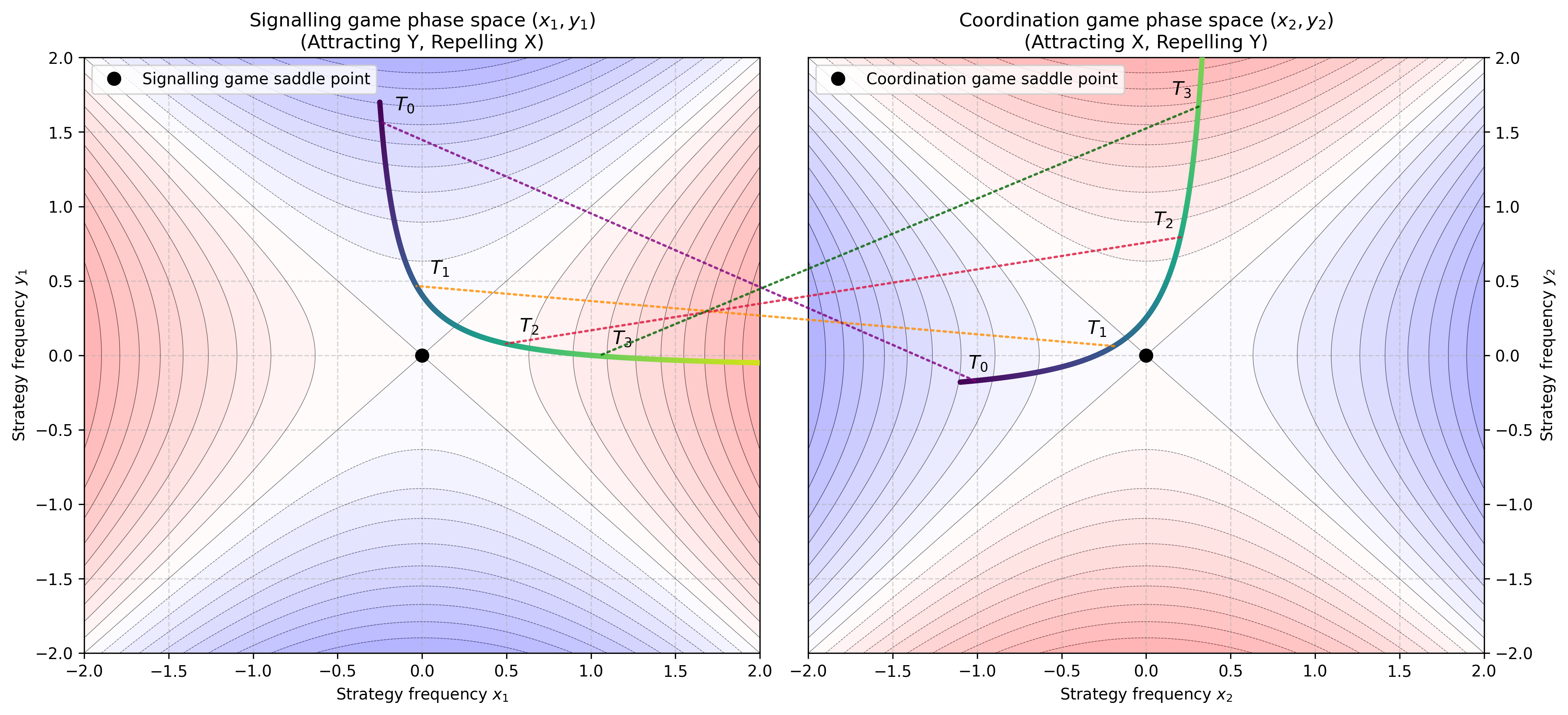}
    \caption{Multichannel signaling-coordination games: synchronized evolutionary trajectory across two saddle landscapes. Left: The trajectory is pulled toward the equilibrium along the $y$-axis: $y$ is the stable (attracting) direction. The trajectory moves away from the saddle in the horizontal direction: $x$ is the unstable (repelling) direction.  Signaling game trajectory initially moves toward the horizontal manifold, then escapes along the $x$-direction. Right: The trajectory is pulled toward the equilibrium along the $x$-axis: $x$ is the stable (attracting) direction. The trajectory moves away from the saddle in the vertical direction: $y$ is the unstable (repelling) direction.  Coordination game trajectory initially moves toward the vertical manifold, then escapes along the $y$-direction. }
    \label{fig:coupled}
\end{figure}

Within this loop, the emerging code and cooperative metabolic networks are hypothesized to not only mutually stabilize one another, but the coupling would need to also give rise to self-modeling. In this process, compressed symbolic representations (that is, DNA information or ``software'') should acquire a functional role in shaping the behavioral patterns and phenotypical traits of the underlying protein ``hardware''.  This bottleneck acts as a filtering process, permitting only those configurations that successfully balance informational stability with functional utility to persist. Such configurations must achieve system-level coherence between genotype and phenotype, satisfy multiple and competing selection pressures (including replication fidelity, catalytic efficiency, and resource constraints), and integrate spatial structuring, selective permeability for metabolite retention, and genetic regulation to suppress freeloading. Failure to evolve and stabilize this coupled set of mechanisms would likely prevent life from advancing beyond rudimentary and easily disrupted proto-cell communities. 
Figure~\ref{fig:coupled} illustrates saddle-type game-theoretic equilibria in multichannel signaling-coordination games where an evolutionary trajectory passes across two saddle landscapes.

From the perspective of algorithmic bottlenecks, the early emergence of widely shared genetic code on Earth does not contradict its difficulty. Rather, it suggests that the coding threshold is characterized by saddle-type equilibria that must be crossed early, before evolutionary dynamics settle into stable non-coding regimes. Trajectories that fail to traverse this bottleneck not only stabilize elsewhere, but may also block access to the bottleneck altogether. For example, early replicator systems may have achieved considerable complexity through catalytic RNA networks without evolving protein-based translation; increasing dependence on such networks could have reduced opportunities for the emergence of a translation apparatus. Thus, this bottleneck is difficult both to enter and to traverse. However, trajectories that successfully cross it may subsequently be drawn toward a stable coding regime, where selection rapidly reinforces and preserves the coding solution. Thus, the genetic code may be both rare across evolutionary possibilities and early in those rare histories where it succeeds.

\subsection{Emergence of symbolic language}

It is well established that, while cooperation can exist without language, the emergence of language substantially enhances and restructures cooperative behavior~\cite{szathmary2015toward}.  Agents participate in collective actions and experience similar environmental feedback, successes and failures. Because agents experience themselves and others acting within the same loop of communication and cooperation, they can eventually form compressed, symbolic representations that model both their own behavior and that of the group. Ultimately, this generates a basis for differentiating between ``self'' and ``non-self'', and ultimately, for self-modeling: ``the key criterion for modern human behavior is not the capacity for symbolic thought but the use of symbolism to mediate behaviour''~\cite{henshilwood2009reading}. These dynamics motivate an examination of the material, behavioral, and cognitive precursors that made such self-modeling and symbol-mediated regulation of behavior possible, and that ultimately culminated in the symbolic explosion.

\subsubsection{Precursors of the symbolic explosion}

Humans stand apart from all other primates by the use of hard, durable tools, mastery of fire, deliberate burial practices, and creation of art~\cite{fletcher1993evolution,coolidge2005working,pettitt2010palaeolithic,scherjon2015burning}. In turn, these practices led to spatial patterning and shape creation, which formed the precursors of verbal language~\cite{fletcher1993evolution}:
\begin{quote}
``The hard debris provided markers on the landscape for the location of human groups, a feature reinforced by the visibility of fires and the marking of the landscape with art. Human activity became entangled with these locational signals, creating an ever more complex milieu for our behaviour and putting great selective pressure on the analytic ability needed to exploit this information. Those humans who could perceive and remember the patterns of markers would have been advantaged in moving around the landscape, whether seeking to meet or avoid other humans.''  
\end{quote} 
In particular, once hominins started using color such as ochre circa 400--320,000 years ago~\cite{gibbons2018complex}, they could mark trails and territory.  Crucially, hard objects, such as stone and bone debris scattered on occupation sites, not only left durable markers on the landscape but also reshaped and complicated our modes of interaction. Locational information became essential for guiding approach and avoidance behaviors, and the palimpsests of accumulated, layered debris that formed over time provided observers with cues about temporal sequence, activity order, and relational context (see Box 4). 

\newpage


{\textbf{Box 4: Expanding cognitive horizons: The emergence of symbolic reference.}}

\boxsection{

The increasing elaboration of stone-tool production demanded a higher cognitive capacity, with ever larger brains capable of remembering and executing progressively longer and more intricate action sequences~\cite{coolidge2005working}. The controlled use of fire generated spatial signals visible over substantial distances, creating reference points that extended far beyond the immediately accessible environment~\cite{scherjon2015burning}. This capacity to determine the location of other hominins at long range --- often without being detected --- offered further selective advantages.

The emergence of abstract symbols may have been supported by an expansion of mental time horizons. The cognitive mechanisms underlying episodic memory and future projection are closely related, and increased capacities for remembering the distant past may have been accompanied by improved abilities to imagine distant futures~\cite{suddendorf2007evolution}. As these temporal horizons expanded, hominins may increasingly have represented events extending beyond their own lifetimes, fostering more abstract conceptions of time analogous to already existing abstractions of space. Such an emerging awareness that the temporal world extends beyond the individual may have found expression in practices such as the deliberate burial of the dead, although the cognitive significance of mortuary behavior remains debated~\cite{pettitt2010palaeolithic}. Together, these developments may have contributed to an emerging view of reality as extending beyond immediate personal experience~\cite{deacon1998symbolic}, while also supporting richer representations of other agents as individuals with perspectives and trajectories independent of oneself.

We note that such capacities are not unique to humans: non-human animals are known to exhibit forms of spatial, temporal, and even social projection without possessing symbolic language. While virtual projection of space, time, and agency can exist without language~\cite{clayton1998episodic,clayton2003can,suddendorf2007evolution,call2008does}, what appears to be distinctive is not abstraction itself~\cite{grush2004emulation,anderson2016comparative}, but the stabilization of abstract representations through shared, conventionalized, and recursively manipulable symbols~\cite{deacon1998symbolic,fitch2010evolution} (see also next subsection).

Art depicting animals, humans, and composite beings that existed independently of the observer emerged around 70–50,000 years ago and became widespread during the Upper Paleolithic: this has often been interpreted as evidence for increasingly abstract representational capacities~\cite{donald1993origins,conard2003palaeolithic}. Such artistic expressions may have deeper roots in the intentional collection and processing of pigments, particularly ochre, documented in Africa during the Middle Pleistocene, dating back 300–400,000 years~\cite{barham2002systematic}. Significantly, art provided a means of storing information outside the brain, thus enabling increasingly sophisticated and structured forms of information processing~\cite{donald1993origins}. 
Such external symbolic representations may also have encouraged increasingly structured relationships among signs, including shapes nested within other shapes, potentially foreshadowing some of the recursive properties characteristic of later symbolic systems~\cite{donald1993origins}.
These developments also encompassed the capacity to reimagine the familiar and to conjure entities never directly observed, for example, the lion-headed human (L\"owenmensch) statuettes dated to 40--30,000 BP~\cite{conard2003palaeolithic}.
}


\

This process can be naturally viewed in terms of niche construction~\cite{trappes2022individualized,harre2024ai}. Once humans began to modify their environment symbolically --- through marks, tools, conventions, or shared representations --- the resulting environment supported the groups that could interpret, manipulate, and extend those symbolic structures. In this way, symbolic practices reshaped the selective landscape, creating a positive feedback loop in which environmental modification and symbolic capacity mutually reinforced one another and led to stabilization of symbolic systems beyond the capacities of non-human animal communication~\cite{deacon1998symbolic,tomasello2005understanding,fitch2010evolution}.

In short, the symbolic explosion was preceded by several interrelated innovations: the use of durable objects as landscape markers and cues (i.e., spatial patterning), increasing sophistication in stone-tool production, the controlled use of fire (including fire-based signaling), a growing cognizance of space and time existing independently of the individual, and the advent of art-making practices such as shape creation and color-coding.

\subsubsection{Least-effort linguistic game}
\label{lelg}

When individuals engage in collective actions, their repeated interactions create a common reference frame with aligned expectations and interpretations, allowing symbols or signals to reliably refer to the same things for different agents. However, this common reference frame comes at a cost. 

A well-developed class of information-theoretic models~\cite{cancho2003least,cancho2005variation,cancho2007global,prokopenko2010phase,salge2015zipf} examines the evolution of language, conceived as a mapping between symbols and objects through the principle of least effort, which seeks to minimize the combined communicative burden of both speaker and listener. These models contrast, for instance, a degenerate system employing a single signal for all objects (thus, minimizing speaker effort) with a fully disambiguated system assigning a unique signal to each object (thus, minimizing listener effort). Different formulations introduce distinct cost functions. In spoken language, for example, the  intrinsic cost of using a signal may be approximated by the time required to articulate the word, which imposes effort on both speaker and listener. 

An early approach by Mandelbrot~\cite{mandelbrot1953informational} treated language as a symbol-production process in which each word carries an inherent cost that increases logarithmically with its complexity. More recent formulations of the least-effort principle propose minimizing a composite cost function that balances the communication inefficiency (measured by the sum of speaker's effort of encoding objects and the listener's effort of decoding signals) with the intrinsic cost of each signal (word)~\cite{salge2015zipf}. Notably, models in this class often exhibit a phase transition separating  degenerate languages that use a single signal and fully disambiguated languages with one-to-one mappings. This transition marks the critical coding threshold at which the competing information-theoretic costs of the speaker and listener come into balance.

Similarly, one may examine the emergence and evolution of language within a game-theoretic framework in which players act as speakers (encoding information, e.g., mapping signals to objects) and listeners (decoding information, interpreting signals). Players interact under payoff structures that reward reliable and efficient communication, reflected in more accurate belief updates, enhanced trust or group inclusion, and penalize ambiguity or excessive costs, such as increased cognitive load or delayed responses. For example, high payoff is achieved when sender and receiver converge on an unambiguous yet low-effort mapping, whereas low payoff results from extreme polysemy (a single signal referring to multiple objects) or high synonymy (multiple signals referring to the same object). Within such models, strategies may range from random signal-object assignments to more structured conventions that minimize miscommunication effort. Evolutionarily stable strategies may emerge when signaling systems settle into equilibria that balance efficiency with cost~\cite{crandall2018cooperating}. In this way, game-theoretic interactions can naturally give rise to increasingly efficient signaling conventions, providing plausible evolutionary precursors to the language threshold.

However, if signaling becomes too noisy or too costly, reliable and rapid recognition, and thus effective communication, breaks down. As with codon-to-amino acid mappings, the space of possible signal-object assignments is combinatorially vast, and interacting players may fail to agree on a common signaling scheme. This can produce fragmented or incompatible systems of communication that favor short-term survival strategies over long-term optimization of the language.

From a dynamical-systems perspective, such fragmented communication systems may correspond to saddle-type coexistence equilibria, in which partially established signaling conventions are stable to some perturbations but unstable to others. In these states, populations can be temporarily drawn toward a common signal-object mapping, creating the appearance of linguistic convention, yet small fluctuations, competing conventions, or stochastic errors may be amplified along unstable directions. Consequently, particular conventions may temporarily dominate, but remain susceptible to displacement by alternative mappings, preventing convergence to a universally shared language. Similar instability and equilibrium-selection phenomena have been studied in evolutionary signaling games and replicator-based models of language evolution~\cite{huttegger2014some}. The resulting dynamics are characterized by continual reorganization around unstable coexistence states, making the long-term stabilization of weakly aligned linguistic systems difficult to achieve.

Just as the early emergence of genetic coding reflects the traversal of a narrow algorithmic bottleneck rather than a long waiting time, the emergence of symbolic language can be understood as a rare arrangement of mutually dependent signaling and coordination requirements. Once a sufficiently robust and locally shared convention that supports communication despite ambiguity and variation is achieved, it rapidly stabilizes and expands. However, if missed, alternative lineages may be confined to stable yet non-symbolic communicative niches, e.g., rich in indexical and iconic signaling, but lacking recursion, abstraction, and collective control mechanisms such as norms.

Under our conjecture, coding thresholds can manifest as saddle-type bottlenecks whose stable manifolds occupy only a narrow region of state space. In this view, phase transitions remain one possible expression of evolutionary hardness, but not its only source.
This perspective can be naturally connected to the optimization framework of Ferrer-i-Cancho~\cite{ferrer2025class}. Rather than requiring languages to remain finely tuned at a critical point, the language threshold may correspond to reaching a restricted class of near-optimal coding solutions. Hardness then reflects the difficulty of evolutionary search and coordination required to enter this class, with criticality emerging as a special case. Importantly, this yields a testable criterion: human languages appear to occupy such near-optimal coding regimes, whereas non-human communication systems do not, suggesting that the language threshold may be defined by entry into a distinctive class of coding-optimal systems~\cite{ferrer2025class}.

This interpretation does not imply that other Earth species, e.g., non-human apes, are intrinsically incapable of language. Rather, it suggests that once a lineage becomes highly adapted around a particular communicative architecture, the evolutionary trajectory leading to a symbolic coding transition may occupy an increasingly narrow region of phase space. Recent discoveries of increasingly structured communication in birds, cetaceans, and other species~\cite{engesser2024seeds,sharma2024contextual,arnon2025whale,phaniraj2026dynamic} are consistent with the saddle-point bottleneck hypothesis: many evolutionary trajectories may approach the language threshold along stable directions, acquiring combinatorial and compositional features. Such trajectories may linger near the threshold and exhibit increasingly language-like properties. Nevertheless, they may fail to traverse the bottleneck, being diverted along unstable directions before achieving recursively closed symbolic language with cumulative symbolic inheritance. Thus, a species' capacity for language is distinct from the likelihood of traversing the language bottleneck through natural evolutionary search.

An interesting boundary case is the waggle dance of honey bees that perform a sophisticated movement to communicate displaced spatial location of food sources, water, or new nest sites to hive mates~\cite{frisch1993dance}.  In particular, dancing bees are able to encode navigational information and translate waggle dances into a vector guiding a searching flight, essentially solving a computational, mapping problem, which in turn is decoded by follower bees~\cite{dyer2002biology}.  However, the waggle dance is not recursively manipulable: it does not support embedding of signals within signals, composition of messages into higher-order structures, or operations such as negating or conditionalizing~\cite{hauser2002faculty,fitch2010evolution}. Moreover, while being symbolic in a minimal sense, the dance leaves no durable external record, and cannot be revisited, edited, or re-interpreted after the performance ends. Even though the waggle dance is socially shared, it does not enable cumulative modification of symbols or trans-generational symbolic accumulation~\cite{deacon1998symbolic}. In short, it can be argued that bees did not cross the symbolic language threshold.

While human spoken language does not generate a permanent material record, we note that it can nonetheless create relatively durable external representations through collective memory, ritualized repetition, teaching, storytelling, and so on. Many complex societies maintained extensive historical, legal and cosmological knowledge before writing~\cite{ong2003orality,vansina1985oral}. Oral traditions, pedagogical lineages, and linguistic conventions function as forms of distributed external memory that can preserve information through repeated transmission across many generations, albeit with higher transmission error than written records, due to drift, loss, and reinterpretation. Writing externalizes symbolic content into a medium that substantially reduces dependence on memory and repeated reconstruction, and should be understood not as the origin of external symbolic storage, but as a dramatic enhancement of its persistence, fidelity, and scalability~\cite{goody1986logic}. Thus, the relevant distinction is not durability \emph{per se} but fidelity of preservation.

\subsubsection{The coordination game}

During collective actions, each agent adjusts their behavior and internal models based not only on their own outcomes, but on patterns that are reinforced through responses of others, enabling coordination and prediction. In particular, the confrontational scavenging scenario suggests that proto-linguistic capacities in \emph{Homo erectus} coevolved with early forms of cooperative behavior among individuals who were not necessarily closely related~\cite{szathmary2015toward}. In this scenario, communication and cooperation must interleave together in a synergistic fashion, as otherwise, the resulting imbalance imposes a significant fitness cost. 

In analyzing this cooperative scenario, Szathm{\'a}ry~\cite{szathmary2015toward} compared two well-known game-theoretic setups. In the $n$-person Prisoner's Dilemma, individuals may choose to cooperate or defect: cooperation entails paying a contribution cost, while defection does not. All contributions are linearly aggregated, scaled by a reward factor $r$, and the resulting sum is evenly redistributed to all individuals, irrespective of their contribution. To defect is known to be the only evolutionary stable strategy (ESS) when $r < n$, which leads to complete absence of cooperation and a Pareto-inefficient equilibrium~\cite{archetti2012game}. This is in stark contrast with the $n$-person public goods game, also known as the Teamwork Dilemma.  

As argued by Szathm{\'a}ry~\cite{szathmary2015toward}, the dynamics of cooperation in the confrontational scavenging scenario follows the Teamwork Dilemma where the collective benefit increases with the number of cooperators in a non-linear (sigmoid) fashion, and hence, attains a cooperative equilibrium. This dynamics is attributed to language which allows for a negotiated division of labor~\cite{szathmary2015toward}: 
\begin{quote}
``This transition happened in early \emph{Homo erectus}, who faced the problem of starvation due to the disappearance of fruits in that period. There was, however, plenty of meat around, including carcasses of the megafauna. ...To use this resource, three crucial actions are needed: First, members of the group who cannot know about the carcass must be informed about its nature, location, and distance; second, they need to be recruited; and, third, execution of the task requires intense cooperation with limited opportunity to cheat. ...It was this niche that allowed a wedge to penetrate the previous animal communication system by signals for displaced items. Given the fact that by then \emph{erectus} had already had a large brain and was very likely equipped with Machiavellian social intelligence (13, 14), the process did not stop there, and protolanguage with increasing richness of symbolic reference started to evolve, to be followed by syntax that presumably emerged with the speciation of \emph{Homo sapiens} (12).''  
\end{quote}

In general, advances in tool use increase the potential payoff of cooperation, while simultaneously raising the complexity of trust formation and signaling. The resulting tension between individual safety and collective gain can also be captured by the \emph{Stag Hunt} game (i.e., the assurance game, trust dilemma, or common-interest game)~\cite{fang2002adaptive}. In this framework, two or more agents choose between non-cooperative strategies of acting alone (safe but yielding low payoffs, analogous to hunting a hare), or cooperative strategies that are risky but potentially highly rewarding (analogous to hunting a stag). Although communal action generates higher payoffs when agents successfully coordinate and combine their skills, it crucially depends on mutual trust.

Although the payoff structure of the Stag Hunt game rewards mutual cooperation and penalizes unilateral risk-taking, achieving the cooperative equilibrium requires trust and mutual assurance. In the absence of trust and under uncertainty about others' actions, the non-cooperative equilibrium becomes \emph{risk-dominant}, while the cooperative (payoff-dominant) equilibrium is not. As a result, cooperation is difficult to sustain (not robust) under uncertainty: players tend to favor the safer, lower-payoff option when facing the possibility of coordination failure. Consequently, the Stag strategy is not evolutionarily robust without dependable coordination mechanisms. Early intelligent agents may oscillate between cooperation and defection, producing easily destabilized collective arrangements. Without stabilizing proto-conventions or proto-norms, populations are likely to retreat to safer, though less collectively optimal, strategies.

From a dynamical-systems viewpoint, the sensitivity of outcomes to variations in trust, signaling and other perturbations can be understood through the presence of a saddle-type equilibrium that separates the basins of attraction of cooperation and non-cooperation. Which equilibrium is risk-dominant depends not on the existence of the saddle itself, but on the relative sizes of the corresponding basins of attraction. Under evolutionary dynamics such as replicator dynamics, mutual cooperation and mutual defection are typically stable attractors, while an intermediate mixed-strategy equilibrium acts as a saddle point. The saddle's stable manifold forms a basin boundary: populations that begin on one side converge toward the payoff-dominant cooperative equilibrium, whereas those on the other side converge toward the risk-dominant non-cooperative equilibrium. Trust, signaling, and emerging social conventions can therefore be viewed as mechanisms that shift populations across this boundary or enlarge the basin of attraction of cooperation~\cite{tria2026language,san2025individual,heinsalu2025language}. In their absence, uncertainty and stochastic fluctuations may drive populations toward the non-cooperative attractor, rendering cooperative arrangements difficult to sustain.

\subsubsection{Algorithmic bottleneck in the emergence of the symbolic language}

The teamwork dilemma (formalized as a public-goods game or the Stag Hunt game) and the least-effort linguistic game can be modeled as coupled through a self-referential feedback loop, in which the output of each serves as an input to the other. The outcome of the linguistic game --- the emergence of a shared symbolic code (language) --- constrains the collective benefits achievable by teams of individuals, via coordination, trust, tool use, and the efficiency of cooperative action. In turn, the payoff structure and outcomes of cooperation feed back into the linguistic game by influencing which signaling conventions and norms are stabilized, refined, or abandoned.

This coupling would induce a multichannel dynamic analogous to that considered in the co-evolution of genetic code and metabolism. On the one hand, an effective and broadly shared language can stabilize cooperation, increase group-level payoffs, and expand the range of viable cooperative strategies and tool use. On the other hand, cooperative success creates selection pressure for signaling systems that are more expressive, reliable, and widely shared, thereby promoting the convergence of linguistic conventions. However, this same coupling can also destabilize otherwise well-behaved dynamics: failures of signaling can undermine trust and cooperation, while breakdowns in cooperation can prevent the stabilization of shared linguistic codes.

In effect, desirable properties of separate games are no longer guaranteed in the coupled system. Instead, the joint (multichannel) dynamics may either converge to mutually reinforcing equilibria, in which language and cooperation stabilize one another, or collapse into regimes characterized by defection, fragmentation of signaling conventions, and the loss of collective benefits. In this sense, the coupled signaling--cooperation system introduces another algorithmic bottleneck. Only those configurations  in which symbolic communication (``software'' or compressed functional information) and cooperative behaviors and tool use (``hardware'') are jointly compatible can persist and evolve further.  Importantly, this loop can be interpreted as involving self-modeling of shared experiences: a body of interactions, perceptions, and outcomes that are jointly encountered by multiple agents and that become mutually recognizable and interpretable over time~\cite{lewis2008convention,tomasello2005understanding,clark2015surfing,heyes2018cognitive}.  
		
The self-referential nature of the ``software''--``hardware'' dynamics is highlighted in the reflection-capable computational architecture described by Barron et al.~\cite{barron2023transitions}:
\begin{quote}
``Natural language also provides an efficient means for externalizing control flow by distributing complexity among groups of individuals, allowing groups to perform tasks too complex for any individual [92]. Dor [24] has proposed that language enables ‘collaborative computation’ between groups of individuals. This crucially involves treating utterances not merely as informative messages but as instructions which can in turn influence control flow. Collaborative work can enable the small-scale apprentice learning that allows for technological transfer and cooperative hunting [93]. On a broader scale, our ability to distribute computational labour allows for complex computations that would be beyond the grasp of any particular human. Reflection thus enables distinctively human forms of learning and cultural accumulation.''
\end{quote}

The emergence of symbolic artifacts, the capacity to store symbols externally, and the evolution of symbolically mediated behavior can be related to a  cognitive change that enabled modern humans to represent different and even conflicting perspectives of objects: ``the capacity to simulate how objects look from conflicting viewpoints''~\cite{henshilwood2009reading}. This also assumes a capacity to represent the perspective of others on objects (including oneself), for example, the practice of wearing personal ornaments like beads indicates an underlying sense of self-awareness and self-recognition~\cite{henshilwood2009reading}. Thus, it has been suggested that the cognitive capacities required to transform an object into a symbol are closely tied to the ability to adopt a self-reflective perspective~\cite{henshilwood2009reading}. 

Crucially, perspective-taking appears vital for crafting formal tools, as it allows the tool-maker to maintain awareness of the subordinate category of the tool during production. For example, the Middle Stone Age artefacts found in Blombos Cave (southern Africa) provided evidence for an early bone industry, with tools falling not only within the general category of ``bone tools'' but also within subordinate categories such as ``bone projectile points'' or ``bone awls''~\cite{henshilwood2001early,henshilwood2009reading}.

The analysis of the precursors of symbolic language   through the cognitive ability to imagine and mentally simulate how objects appear from multiple, potentially divergent viewpoints, offered by Henshilwood and Dubreuil~\cite{henshilwood2009reading},  supports our thesis on the ``algorithmic origin of language''. Specifically, we can identify the ``symbolic explosion'' or the ``language threshold'' with the emergence of a tangled information hierarchy between language and symbolically mediated behavior. Importantly, the self-reflecting ability, enabled by this hierarchy, allowed for recursive syntax, with clauses embedded within clauses \cite{jackendoff1999possible,hauser2002faculty}: ``the Rubicon of modern language probably lies in recursion'', as without recursive syntax, one cannot express or integrate conflicting perspectives~\cite{henshilwood2009reading}.

In summary, crossing the ``language threshold'' and associated coordination barrier --- amplified by strategic uncertainty and environmental risks --- poses a central challenge for cooperating tool-using agents. To exploit high-reward opportunities such as hunting large prey, building tools, or knowledge sharing, these agents must reliably communicate and coordinate. This coupling constitutes an algorithmic bottleneck that constrains the transition to symbolic language, whose successful scaling alongside tool use and cooperative behaviors is not guaranteed.

\section{Discussion}

In this article we explored the conjecture that the hardest steps on the path towards  technological societies capable of interstellar travel can be interpreted as algorithmic bottlenecks, shaped by (nested) tangled information hierarchies. Symbolic encoding, interpretive machinery and recursive closure may be viewed as prerequisites for these bottlenecks, whereas durable external symbolic memory and recursive symbol manipulation are the informational capabilities that fully emerge once those prerequisites are established. We conjecture that the mechanisms generating these bottlenecks may correspond to saddle-type equilibria in evolutionary game-theoretic models, which reduce the probability of complex lifeforms and societies. 

We point out that both considered evolutionary transitions --- the emergence of genetic code and the emergence of symbolic language --- combine sophisticated signaling and strategic coordination problems in competitive and noisy environments. These dynamics must not only stabilize themselves, but also establish new information architectures that unlock further evolutionary phase spaces, enabling novel ways to store, transmit and interpret information.  Therefore, these evolutionary transitions are far from guaranteed. Their success depends on navigating biochemical and socio-biological landscapes in which signaling and coordination remain vulnerable to perturbations that push dynamics along unstable directions, and where defection, whether through reduced signaling effort or exploitation of shared resources, is tempting. Societies with advanced technologies may therefore be rare because at least one code-invention is very hard, path-dependent, or successful only along a narrow range of evolutionary trajectories. Thus, the Great Filter arising from algorithmic bottlenecks may leave only a small subset of trajectories capable of reaching long-term survival. 

To reiterate, a hard algorithmic bottleneck need not be crossed late. It may instead be traversed early in those rare evolutionary histories where the requisite conditions arise. Furthermore, because the second bottleneck lies within the evolutionary landscape made accessible by the first, the two transitions should be regarded as sequentially dependent (nested). Thus, the fact that  both the genetic-coding and symbolic-communication thresholds have been crossed on Earth does not imply that such transitions are generally easy. Rather, it suggests that our evolutionary trajectory was among the few that reached and traversed both bottlenecks before becoming trapped in alternative stable regimes. Furthermore, these thresholds should not be understood as instantaneous inventions emerging without antecedents. Most likely, they have been preceded by periods of gradual evolutionary accumulation, leading to a discontinuous expansion of the accessible evolutionary phase space.

A recent study hypothesized the existence of a biological arrow of time underlying major evolutionary transitions, and formalized this conjecture through the lens of computation theory~\cite{prokopenko2025biological}. According to this interpretation, biological organisms are hierarchical dynamical systems that generate regularities in their phase-spaces through interactions with their environment. These emergent information patterns can be encoded  within the organism (e.g., via the genotype-pheno\-type mapping), leading to ``fundamental tensions (computational inconsistencies) between what is encodable at a particular evolutionary stage and what is potentially realizable in the environment''~\cite{prokopenko2025biological}. This study argued that open-ended evolution comprises a continual process in which a resolution of these tensions drives an evolutionary transition, expanding the problem-space and then generating new tensions in the expanded space. 

Computation-theoretically, the problem-space expands when computational novelty (e.g., a new ``code of life'') is generated by self-modelling agents which gain access to meta-simulation~\cite{prokopenko2025biological}, e.g., by Turing oracles or recursively generated logics~\cite{turing1939systems,post_recursively_1944,soare2009turing,markose2004novelty,markose_complex_2017}, or receive the feedback from external environment, e.g., via natural selection~\cite{svahn2023ansatz}. Self-modeling, e.g., via genome or language, yields big advantages: robustness, division of labor, and cooperation at scale.  Under this perspective, crossing the ``coding threshold'' may require agents to effectively  ``Jump Out Of The System'' (JOOTS)~\cite{hofstadter_go_1980}, only to inevitably encounter yet another incomputable paradox within the expanded, higher-level system.

The two algorithmic bottlenecks examined in this work punctuate the path of open-ended evolution by necessitating the emergence and stabilization of a ``software''--``hardware'' separation under selection pressures. Crossing the associated coding thresholds requires the formation of self-referential, self-modeling loops that, in turn, enable novel inheritance systems and shared representational frameworks.

This perspective suggests that despite continually opening new evolutionary possibilities, this process may progressively narrow the evolutionary path available for recursively closed code--interpreter systems. Every successful coding transition expands the accessible phase space while simultaneously increasing its topological complexity. The resulting accumulation of nested information hierarchies may lead to a diminishing set of high-dimensional saddle-point pathways connecting viable regions, thereby reducing the fraction of traversable evolutionary trajectories. This may provide a mechanism by which open-ended evolution generates increasingly severe algorithmic bottlenecks that constrain subsequent innovations.

Contemporary evolutionary AI provides an instructive analogue of these coding thresholds. For example, neuroevolution studies have shown that strong attentional, memory, and information-processing bottlenecks can drive agents to develop compressed, robust, and partially self-interpretable internal representations~\cite{tang2020neuroevolution}. Such results suggest that informational constraints need not merely limit adaptation; they may also catalyze the emergence of higher-level abstractions that improve coordination between representation and action. In this sense, artificial evolutionary systems offer a computational proof-of-concept for our conjecture that algorithmic bottlenecks can promote the formation of new information architectures rather than simply impede evolutionary progress.

As has been noted, ``just as the evolution of powerful epigenetic inheritance systems allowed the evolution of complex multicellularity, natural language allowed the emergence of complex human societies''~\cite{jablonka1994inheritance,szathmary2015toward}. However, as the capacities for communication and shared representation scale into forms of collective intelligence, societies begin to confront analogous ``coding thresholds'' at the cultural and institutional level. These higher-order systems would need to stabilize cooperative codes, shared protocols, norms, and governance mechanisms in environments where strategic pressures can easily destabilize them. In this way, the challenges faced by evolving technological societies echo the earlier algorithmic bottlenecks of biological evolution.

The emergence of technological society can be modeled as an iterated or evolutionary $n$-player \emph{Prisoner's Dilemma} (Commons Dilemma), in which cooperation must be maintained despite incentives to defect. As societies become increasingly interconnected and globalized, the \emph{tragedy of the commons} extends across planetary boundaries, amplifying the difficulty of attaining cooperative equilibria. National governments and legal systems struggle to construct multilateral institutions that are robust enough to manage shared resources, regulate collective \emph{externalities} (e.g., climate change, biodiversity loss, pollution), and coordinate actions across diverse actors whose interests are only partially aligned~\cite{wiedmann2018environmental}. 

Another generic aspect of the challenges facing complex societies is \emph{short-termism}: a decision-making bias prioritizing short-term rewards over long-term outcomes, which can manifest as a lack of inter-generational equity, e.g., with respect to the climate change crisis~\cite{lee2023ipcc}. For example, there is a strong argument that our modern society (the Anthropocene) is operating at --- or has surpassed --- the Earth’s carrying capacity in respect of planetary boundaries~\cite{rockstrom2024planetary}. Without overarching constraints, short-termism is unable to capture complex responses required by long-term challenges~\cite{fisher2023long}. 

Strategies that may be employed by technological societies in an attempt to resolve these challenges include innovation (in creating a new shared code of societal life). Crucially, innovation may go beyond the directed invention of a specific function, and instead recombine the accumulated material features and devices that were often generated for unrelated purposes. As emphasized by Fletcher~\cite{fletcher1995limits}, novel capacities frequently arise when existing assemblages of tools, practices, and infrastructures are brought into new configurations, thereby acquiring functions not anticipated at their origin\footnote{For example, redesigning pendulum clocks into marine chronometers solved the ``longitude problem'' and revolutionized maritime navigation.}. Societies that encourage experimentation and variation are more likely to introduce new modes of information processing, coordination, and control, leading to qualitative shifts in collective behavior rather than incremental improvement.
Other strategies, such as steady-state dynamics (e.g., steady state economy~\cite{daly1977steady}), or degrowth/post-growth, are aimed at reducing production, consumption and environmental impacts, but these approaches have their own challenges~\cite{keysser20211,kallis2025post}. Overall, technological societies that fail to build resilient institutions or internalize all key impacts of technological development, such as environmental or social costs, may decline or collapse before reaching interstellar/astroengineering capability.

Another emerging source of systemic risk arises from advanced forms of Artificial Intelligence (AI), such as agentic AI~\cite{evans2026agentic}. As has been observed, human society can be understood as a kind of ``operating system'': a multi-layered structure built from stories, norms, legal frameworks, monetary systems, and religious worldviews, all of which are mediated by language~\cite{harari2023yuval}.  Harari~\cite{harari2023yuval} argued that generative AI systems have effectively ``hacked'' this societal operating system, intervening directly in the linguistic substrate on which collective understanding and decision-making depend. In other words, AI has gained the capacity to act upon our societal ``code of life'', influencing the symbols, narratives, and discourses through which large-scale cooperation is organized. The result is a destabilizing feedback loop: AI systems trained on human discourse begin to influence that same discourse, altering the informational ecology in ways that may erode cooperative norms or amplify collective vulnerabilities. 

This ``operating system'' metaphor again underscores the algorithmic character of tangled societal hierarchy in which language-mediated information functions as system software coordinating both the social substrate (hardware) and its various informational resources (other software components).  
It is well-known that some of the most destructive systemic threats arise from exploiting vulnerabilities in a system's capacity for accurate self-description~\cite{markose2021genomic}. Examples include computer viruses that alter host code to invert the intended computation~\cite{marion2012turing}, and novel antigens that compromise the immune system's ``self versus non-self'' discrimination, precipitating autoimmune diseases~\cite{geenen2021thymus}. By the same logic, the ``hacking'' of language-mediated societal functions represents an existential risk, because it targets the very substrate through which collective coordination is maintained. Thus, the formation of a stable tangled societal hierarchy may constitute yet another algorithmic bottleneck, susceptible to disruption by accelerating, and potentially runaway, AI-dominated dynamics. A precise delineation of the emerging ``coding threshold'' for societal intelligence~\cite{kim2026reasoning,evans2026agentic} remains unresolved, representing a future hard step that humanity has yet to navigate.

\section{Conclusion}

Informally, life, mind, and society are all information engines that extract order from disorder. However, early conditions constrain future possibilities, and Great Filters may correspond to algorithmic bottlenecks, manifested in many ways, as limits in error correction, e.g., mutation rates~\cite{eigen1971selforganization}; cognitive constraints, e.g., Dunbar’s number~\cite{dunbar1992neocortex,harre2016social}; social fragility, e.g., loss of knowledge in collapse~\cite{bostrom2019vulnerable,manheim2020fragile}, and so on. Under our conjecture, each algorithmic bottleneck, beginning with the emergence of genetic coding and followed by the emergence of symbolic language, moves evolution toward the next level of informational complexity. These bottlenecks are nested in the sense that each higher-level transition presupposes the successful stabilization of earlier coding system(s), constraining subsequent evolution to occur only within lineages that have already crossed preceding threshold(s). Specifically, agents capable of symbolic language evolve within lineages that have first stabilized genetic coding, rather than emerge along trajectories that bypass this transition, for example, following purely chemical pathways. We hypothesize that because these tangled information hierarchies are nested rather than independent, the sequence of coding thresholds forms a Great Algorithmic Filter, with generic fundamental properties.  

A limitation of this perspective is the potential inflation of evolutionary filters through anthropocentric biases or generalization from the single known history of life on Earth. Features that appear uniquely important when viewed through the lens of human evolution may reflect path-dependent contingencies rather than general principles or universal evolutionary constraints. Consequently, some elements of the identified bottlenecks may prove less restrictive in alternative biological, cognitive, or ecological contexts, including those realized by other Earth-originating or extraterrestrial intelligences.

Our game-theoretic analysis is intended to shift the interpretation of ``hard steps'' from elusive improbability to equilibrium structure, arguing that algorithmic bottlenecks may correspond to saddle-type equilibria in multichannel signaling--coor\-dination games where signaling, coordination, and control must tightly align. Importantly, this game-theoretic analysis is not intended to provide a detailed mechanistic simulation of the evolutionary dynamics leading up to these bottlenecks. Instead, it aims to diagnose the structural difficulty of these transitions and expose the equilibrium-level constraints that make their traversal inherently difficult. Rather than presuming that complexity inevitably accumulates through adaptive innovations, this hypothesis suggests that information-processing innovations may be vulnerable to strategic collapse due to multichannel failures, thus increasing the risk of systemic fragility and evolutionary dead ends.

\appendix

\section*{Appendix A: Conceptual analogy between genetic and linguistic coding systems}
\label{concept}

The conceptual analogy between genetic and linguistic coding systems is more than merely metaphorical, see Table~\ref{tab1}. In both cases, symbols are inert in the absence of an interpreting machinery capable of mapping them onto functional outcomes. The code itself is arbitrary but conventional --- widely shared within a population --- and therefore requires collective stabilization to persist. Importantly, functional consequences arise only through rule-governed translation, rather than via direct causal interaction between symbols and outcomes. This shared structure suggests that both transitions constitute genuine hard steps in an algorithmic sense, marked by the emergence of a new level of symbolic representation and interpretation within a recursively closed architecture. 

This analogy does not imply a strict or unique correspondence, and alternative mappings are possible~\cite{searls2002language,faltynek2019bases}. Moreover, some of the proposed correspondences remain debated. For example, an individual genome may be likened to a single text (e.g., book), whereas a pangenome may be compared to a corpus, text collection or library.  At a more abstract level, however, genome may also be viewed as the repertoire from which potential texts, utterances, and actions can be generated. In this, generative sense, an individual genome is analogous to linguistic competence~\cite{pattee2001physics,deacon2011incomplete}, whereas a pangenome would be more closely comparable to language or discourse traditions~\cite{boyd1988culture,mufwene2008language}. Another non-unique correspondence involves recursive syntax, which has been compared both to nested RNA secondary structures~\cite{searls2002language,hauser2002faculty}, and to hierarchically organized gene regulatory networks~\cite{davidson2006regulatory}.

In this article, we distinguish symbolic reference from syntax. Symbolic reference denotes the use of socially shared and conventionalized symbols whose meanings are not fixed by immediate context, resemblance, or physical causation. Syntax denotes the combinatorial principles governing how such symbols may be assembled into higher-order expressions, including recursively structured constructions. We remain agnostic regarding the precise historical order of their emergence. Some accounts propose increasingly rich symbolic reference preceding syntax, whereas others argue that the two capacities co-evolved. Our language threshold refers not to either capacity in isolation, but to the stabilization of a jointly interpretable symbolic system capable of supporting open-ended communication, cumulative inter-generational transmission, and shared control flow. An operational perspective treats symbolic reference as a non-degenerate mapping between signals and referents, while syntax emerges when combinatorial organization among signals becomes advantageous under communicative constraints. Under this view, symbolic reference and syntactic structure may arise jointly as consequences of pressures toward efficient communication~\cite{i2005consequences}.

\section*{Appendix B: a glossary of terms}
\label{app}

\textbf{Collectively catalytic system} is a system where multiple components work interdependently to accelerate processes. A collectively autocatalytic set is a collection of molecular entities in which each member's creation is catalyzed by another entity within the same set. As a whole, the set sustains and accelerates its own production~\cite{kauffman1986autocatalytic}.

\textbf{Evolutionary singularities} are defined by De Duve~\cite{de2005singularities} as ``events or properties that have the
quality of singleness, uniqueness``. In other words, evolutionary singularities appear to have occurred once (or extremely rarely) and have left no independent alternatives or parallel realizations. In this sense, an evolutionary singularity is a path-dependent turning point, constraining future possibilities, by establishing a new organizational principle or informational architecture that becomes foundational for all subsequent evolution. Canonical examples considered by De Duve~\cite{de2005singularities} in the context of the Earth's evolutionary record, include the origin of life, the genetic code, and the transition to complex cellular organization, each of which appears universally shared across extant life and admits no known competing realizations.

\textbf{Externality}, in economic theory, refers to a cost or benefit that arises from an agent's activity and affects third parties who are not directly involved in the transaction~\cite{boudreaux2019externality}. For example, motor vehicle emissions is a negative externality, with the societal costs of air pollution --- such as public health impacts and environmental degradation --- are not fully borne by either the producers or consumers of motorized transport. It has been argued that since these third-party effects --- whether negative (external costs) or positive (external benefits) --- are not captured by market prices, they yield potential inefficiencies in resource allocation~\cite{sandholm2005negative,malik2024polarizing}.

\textbf{Iconic references} are signals that resemble or mimic their referents through structural or perceptual similarity. Their meaning arises from likeness rather than arbitrary convention, e.g., gestures, or words that phonetically imitate, resemble, or suggest the sounds they describe (onomatopoeia).

\textbf{Indexical references} are linguistic expressions whose reference shifts depending on the context of the utterance, such as the identity, location, or time of the speaker. Examples include words such as ``I'', ``here'', ``now'', which do not have a fixed, stable referent across different situations. In other words, the interpretation of indexical references is fixed by context, not by convention.


\begin{table}[H]
\centering
\caption{Conceptual mapping between genetic and linguistic coding systems.}
\label{tab:genetic_language_mapping}
\renewcommand{\arraystretch}{1.5}
\begin{tabular}{|p{3.5cm}|p{4.25cm}|p{4.75cm}|p{2cm}|}
\hline
\textbf{Genetic system} & \textbf{Linguistic system} & \textbf{Generic role} & \textbf{References} \\
\hline
Nucleotide bases & Graphemes, letters & Basic symbolic units & \cite{pattee2012does,jakobson1971selected,smith2000concept,barbieri2003organic} \\
\hline 
DNA sequences & Symbol strings  & Abstract, discrete information carriers & \cite{pattee2012does,jakobson1971selected,smith2000concept,barbieri2003organic} \\
\hline
Codons & Words, morphemes & Basic symbolic units combined to form higher-level structures & \cite{pattee2012does,jakobson1971selected,smith2000concept,barbieri2003organic} \\
\hline
Genes & Sentences, lexical items & Minimal functional units carrying meaning or instruction & \cite{pattee2012does,jakobson1971selected,smith2000concept,barbieri2003organic} \\
\hline
Genetic code (codon $\to$ amino acid mapping) & Lexicon, semantic conventions (symbol $\to$ meaning mapping) & Arbitrary but shared mapping between symbols and meanings & \cite{pattee2012does,jakobson1971selected,yockey1992information,smith2000concept,barbieri2003organic} \\
\hline
Translation machinery (ribosome, tRNA, enzymes) & Interpreter (parser plus semantic engine) & Mechanisms that interpret symbolic strings into functional outcomes & \cite{pattee2012does,smith2000concept,barbieri2003organic,deacon2011incomplete} \\
\hline
Ribosome & Parser applying grammar, syntax & Rule-based structure governing valid combinations & \cite{pattee2012does,smith2000concept,barbieri2003organic,deacon2011incomplete} \\
\hline
tRNA adapters & Semantic interpreters, conceptual schemas & Mediators linking symbols to functional meanings & \cite{pattee2012does,smith2000concept,barbieri2003organic,deacon2011incomplete} \\
\hline
Nested RNA secondary structures & Recursive syntax & Nested, recursive symbolic structures & \cite{searls2002language,hauser2002faculty} \\
\hline
Proteins & Communicative acts, structured thoughts and behaviors & Functional products of interpretation & \cite{pattee2012does,pattee2001physics,barbieri2003organic,barbieri2008biosemiotics,deacon2011incomplete} \\
\hline
Protein folding & Pragmatics, contextual interpretation & Context-dependent shaping of function or meaning & \cite{hoffmeyer1997signs,pattee2001physics,barbieri2003organic,deacon2011incomplete} \\
\hline
Gene regulation & Discourse rules, conversational norms & Control of when and how expressions are produced & \cite{hoffmeyer1997signs,pattee2001physics,barbieri2003organic,deacon2011incomplete} \\
\hline
Replication fidelity & Transmission fidelity (learning, teaching) & Accuracy of inheritance across generations & \cite{dawkins1976selfish,boyd1988culture,szathmary1995major} \\
\hline
Mutations & Linguistic variation, semantic drift & Source of novelty and variation & \cite{dawkins1976selfish,boyd1988culture,szathmary1995major,croft2000explaining,croft2008evolutionary,mufwene2008language} \\
\hline
Horizontal gene transfer & Language contact, borrowing & Exchange of information across lineages or groups & \cite{croft2000explaining,gogarten2002prokaryotic,doolittle2007pattern} \\
\hline
Genome & Book / Linguistic competence  & Particular repository of symbolic information & \cite{pattee2001physics,searls2002language,deacon2011incomplete}  \\
\hline
Pangenome & Library, corpus / Language & Collective repository of symbolic information, durable external memory & \cite{boyd1988culture,searls2002language,mufwene2008language}  \\
\hline
\end{tabular}
\label{tab1}
\end{table}


\newpage

\textbf{Iterated Prisoner's dilemma} is a game where two players repeatedly choose to ``cooperate'' or ``defect'', allowing for strategies based on past moves, unlike the single-shot version where defecting is always best. In the iterated Prisoner's dilemma, developing trust and reciprocating actions (like ``Tit for Tat'': start by cooperating and then mirror the opponent's previous move, i.e., cooperate if they cooperated, defect if they defected) can lead to mutual cooperation and better overall outcomes than always defecting.  However, if players know the game will end after a fixed, finite number of rounds, this logic unravels through backward induction. Players will defect in the last round, which means they should  defect in the second-to-last round, and so on, leading to defection from the start. Under these conditions, the cooperative outcome is less robust to uncertainty than the non-cooperative which makes the latter the \emph{risk dominant equilibrium}.

\textbf{Misincorporation} is an addition of a wrong amino acid into a growing polypeptide chain during translation, relative to the codon specified by the mRNA.

\textbf{Mixed-strategy Nash equilibrium} in game theory is a strategy profile where players randomly choose between their available pure strategies with specific probabilities (i.e., there is a probability distribution over different strategies, e.g., $1/3$ assigned to each strategy in Rock-Paper-Scissors game). 

\textbf{Multichannel games} in game theory refer to a framework where multiple, distinct strategic interactions are interconnected, such that the actions, outcomes, or payoffs of one game influence the parameters or choices in another~\cite{donahue2020evolving}.

\textbf{Pareto efficiency} describes a resource allocation outcome without an alternative allocation that increases the payoff of any player without decreasing the payoff of at least one other player.  A Nash equilibrium is payoff dominant if it is Pareto superior to all other \emph{Nash equilibria} in the game.

\textbf{Risk dominant equilibrium} in game theory is a stable Nash equilibrium that has the largest basin of attraction, making it less sensitive to uncertainty. 
That is, players are more likely to choose this profile when they are uncertain about their opponents' actions, favoring strategies that minimize potential losses rather than maximize potential gains (payoff dominance). For example, the Stag Hunt is an assurance game with two pure strategy Nash equilibria: a cooperative, payoff-dominant equilibrium (both hunt stag) and a safer, non-cooperative equilibrium (both hunt hare). While both equilibria are stable, the cooperative equilibrium is not risk-dominant: it is not robust to uncertainty. A lack of trust and the fear of the worst-case scenario makes players choose the safer, less rewarding option, creating a dilemma between the high-reward, high-risk cooperative outcome and the low-reward, low-risk non-cooperative one.

\textbf{Self-regarding signals} in language theory are communicative acts whose reference is anchored to the signaler's own immediate state, perspective, or situation, rather than to shared, abstract, or displaced entities. They convey information about ``me, here, now'', and do not require a shared symbolic code.

\textbf{Short-termism} is a decision-making bias which gives priority to short-term profit or reward at the expense of long-term results and far-seeing action, thus motivating quickly executed projects~\cite{fisher2019perils,wiki-ST}.  Abstract reasoning about time and the effects of decisions and actions assumes the abilities to conceive, represent and imagine (or simulate) dynamic trajectories. These abilities are likely to have evolved in many extraterrestrial lifeforms via adaptations which improved the accuracy of prediction (forward simulation) and explanation (backward simulation), and thus, provided selective advantages.  In turn, a partial inability to look beyond the short-term horizon is a bias that can be generalized to the extraterrestrial lifeforms which developed a concept of time, as well as a capacity to reason about past and future scenarios.

\textbf{Stable equilibrium}, in the context of replicator dynamics, is a population strategy distribution that persists under the evolutionary dynamics: following small perturbations, selection pressures drive the population back toward the equilibrium, rather than away from it, because any mutant or deviating strategy earns lower payoff than the resident mix. A stable equilibrium under replicator dynamics may not be risk-dominant.

\textbf{Technosignature-bearing society} refers to an extraterrestrial society that produces measurable technosignatures: observable properties or effects providing scientific evidence of past or present technology, e.g., radio emissions or large-scale astroengineering~\cite{lingam2021life}.

\textbf{Tragedy of the commons} describes how individuals, acting in their own self-interest, will overuse and deplete a shared, limited resource (a ``common'') even when it is not in anyone's longer-term best interest, leading to collective ruin, exemplified by overfishing or deforestation. This may be exacerbated by \emph{short-termism}.

\textbf{Unstable equilibrium} in game theory is a strategy profile where any slight player deviation pushes the system away from that point, rather than back towards it. For example, in the driving (``chicken'' or the Hawk-Dove game) game, with two drivers moving in opposite directions and choosing to swerve either to the left or to the right of the road, the mixed-strategy equilibrium ($1/2$ probability of each action) is unstable. If a player changes their probabilities, while maintaining the expected payoff under the assumption that the other player's mixed strategy is still ($1/2$, $1/2$), then the other player immediately has a better pure strategy at either (0, 1) or (1, 0).  

In the Rock-Paper-Scissors game, the mixed Nash equilibrium ($1/3$,$1/3$,$1/3$) is a neutrally stable center under idealized replicator dynamics. However, because it is not an evolutionarily stable strategy and not an attractor, arbitrarily small perturbations prevent convergence, leading populations to continually adjust their strategies and cycle indefinitely rather than settle at the equilibrium.

\

\textbf{AI Acknowledgement Statement.}
During manuscript preparation, the Authors used Microsoft Copilot to paraphrase text and improve sentence structure. We confirm that where text was modified by generative AI, the content was reviewed for possible errors, inaccuracies, and bias.


\begin{thebibliography}{167}
\expandafter\ifx\csname natexlab\endcsname\relax\def\natexlab#1{#1}\fi
\providecommand{\url}[1]{\texttt{#1}}
\providecommand{\href}[2]{#2}
\providecommand{\path}[1]{#1}
\providecommand{\DOIprefix}{doi:}
\providecommand{\ArXivprefix}{arXiv:}
\providecommand{\URLprefix}{URL: }
\providecommand{\Pubmedprefix}{pmid:}
\providecommand{\doi}[1]{\href{http://dx.doi.org/#1}{\path{#1}}}
\providecommand{\Pubmed}[1]{\href{pmid:#1}{\path{#1}}}
\providecommand{\bibinfo}[2]{#2}
\ifx\xfnm\relax \def\xfnm[#1]{\unskip,\space#1}\fi
\bibitem[{Anderson(2016)}]{anderson2016comparative}
\bibinfo{author}{Anderson, J.R.}, \bibinfo{year}{2016}.
\newblock \bibinfo{title}{Comparative thanatology}.
\newblock \bibinfo{journal}{Current Biology} \bibinfo{volume}{26},
  \bibinfo{pages}{R553--R556}.
\bibitem[{Archetti and Scheuring(2012)}]{archetti2012game}
\bibinfo{author}{Archetti, M.}, \bibinfo{author}{Scheuring, I.},
  \bibinfo{year}{2012}.
\newblock \bibinfo{title}{Game theory of public goods in one-shot social
  dilemmas without assortment}.
\newblock \bibinfo{journal}{Journal of Theoretical Biology}
  \bibinfo{volume}{299}, \bibinfo{pages}{9--20}.
\bibitem[{Arnon et~al.(2025)Arnon, Kirby, Allen, Garrigue, Carroll and
  Garland}]{arnon2025whale}
\bibinfo{author}{Arnon, I.}, \bibinfo{author}{Kirby, S.},
  \bibinfo{author}{Allen, J.A.}, \bibinfo{author}{Garrigue, C.},
  \bibinfo{author}{Carroll, E.L.}, \bibinfo{author}{Garland, E.C.},
  \bibinfo{year}{2025}.
\newblock \bibinfo{title}{Whale song shows language-like statistical
  structure}.
\newblock \bibinfo{journal}{Science} \bibinfo{volume}{387},
  \bibinfo{pages}{649--653}.
\bibitem[{Axelrod and Hamilton(1981)}]{axelrod1981evolution}
\bibinfo{author}{Axelrod, R.}, \bibinfo{author}{Hamilton, W.D.},
  \bibinfo{year}{1981}.
\newblock \bibinfo{title}{The evolution of cooperation}.
\newblock \bibinfo{journal}{Science} \bibinfo{volume}{211},
  \bibinfo{pages}{1390--1396}.
\bibitem[{Barbieri(2003)}]{barbieri2003organic}
\bibinfo{author}{Barbieri, M.}, \bibinfo{year}{2003}.
\newblock \bibinfo{title}{The organic codes: an introduction to semantic
  biology}.
\newblock \bibinfo{publisher}{Cambridge University Press}.
\bibitem[{Barbieri(2008a)}]{barbieri2008biosemiotics}
\bibinfo{author}{Barbieri, M.}, \bibinfo{year}{2008}a.
\newblock \bibinfo{title}{Biosemiotics: a new understanding of life}.
\newblock \bibinfo{journal}{Naturwissenschaften} \bibinfo{volume}{95},
  \bibinfo{pages}{577--599}.
\bibitem[{Barbieri(2008b)}]{barbieri2008mechanisms}
\bibinfo{author}{Barbieri, M.}, \bibinfo{year}{2008}b.
\newblock \bibinfo{title}{The mechanisms of evolution: natural selection and
  natural conventions}, in: \bibinfo{booktitle}{The Codes of Life: The Rules of
  Macroevolution}. \bibinfo{publisher}{Springer}, pp. \bibinfo{pages}{15--35}.
\bibitem[{Barbieri(2019)}]{barbieri2019general}
\bibinfo{author}{Barbieri, M.}, \bibinfo{year}{2019}.
\newblock \bibinfo{title}{A general model on the origin of biological codes}.
\newblock \bibinfo{journal}{Biosystems} \bibinfo{volume}{181},
  \bibinfo{pages}{11--19}.
\bibitem[{Barham(2002)}]{barham2002systematic}
\bibinfo{author}{Barham, L.}, \bibinfo{year}{2002}.
\newblock \bibinfo{title}{Systematic pigment use in the middle pleistocene of
  south-central africa}.
\newblock \bibinfo{journal}{Current Anthropology} \bibinfo{volume}{43},
  \bibinfo{pages}{181--190}.
\bibitem[{Barron et~al.(2023)Barron, Halina and Klein}]{barron2023transitions}
\bibinfo{author}{Barron, A.B.}, \bibinfo{author}{Halina, M.},
  \bibinfo{author}{Klein, C.}, \bibinfo{year}{2023}.
\newblock \bibinfo{title}{Transitions in cognitive evolution}.
\newblock \bibinfo{journal}{Proceedings of the Royal Society B}
  \bibinfo{volume}{290}, \bibinfo{pages}{20230671}.
\bibitem[{Bostrom(2019)}]{bostrom2019vulnerable}
\bibinfo{author}{Bostrom, N.}, \bibinfo{year}{2019}.
\newblock \bibinfo{title}{The vulnerable world hypothesis}.
\newblock \bibinfo{journal}{Global Policy} \bibinfo{volume}{10},
  \bibinfo{pages}{455--476}.
\bibitem[{Boudreaux and Meiners(2019)}]{boudreaux2019externality}
\bibinfo{author}{Boudreaux, D.J.}, \bibinfo{author}{Meiners, R.},
  \bibinfo{year}{2019}.
\newblock \bibinfo{title}{Externality: Origins and classifications}.
\newblock \bibinfo{journal}{Natural Resources Journal} \bibinfo{volume}{59},
  \bibinfo{pages}{1--34}.
\bibitem[{Boyd and Richerson(1988)}]{boyd1988culture}
\bibinfo{author}{Boyd, R.}, \bibinfo{author}{Richerson, P.J.},
  \bibinfo{year}{1988}.
\newblock \bibinfo{title}{Culture and the evolutionary process}.
\newblock \bibinfo{publisher}{University of Chicago press}.
\bibitem[{Call and Tomasello(2008)}]{call2008does}
\bibinfo{author}{Call, J.}, \bibinfo{author}{Tomasello, M.},
  \bibinfo{year}{2008}.
\newblock \bibinfo{title}{Does the chimpanzee have a theory of mind? 30 years
  later}.
\newblock \bibinfo{journal}{Trends in Cognitive Sciences} \bibinfo{volume}{12},
  \bibinfo{pages}{187--192}.
\bibitem[{Carter(1983)}]{carter1983anthropic}
\bibinfo{author}{Carter, B.}, \bibinfo{year}{1983}.
\newblock \bibinfo{title}{The anthropic principle and its implications for
  biological evolution}.
\newblock \bibinfo{journal}{Philosophical Transactions of the Royal Society of
  London. Series A, Mathematical and Physical Sciences} \bibinfo{volume}{310},
  \bibinfo{pages}{347--363}.
\bibitem[{Clark(2015)}]{clark2015surfing}
\bibinfo{author}{Clark, A.}, \bibinfo{year}{2015}.
\newblock \bibinfo{title}{Surfing uncertainty: Prediction, action, and the
  embodied mind}.
\newblock \bibinfo{publisher}{Oxford University Press}.
\bibitem[{Clark and Chalmers(1998)}]{clark1998extended}
\bibinfo{author}{Clark, A.}, \bibinfo{author}{Chalmers, D.},
  \bibinfo{year}{1998}.
\newblock \bibinfo{title}{The extended mind}.
\newblock \bibinfo{journal}{Analysis} \bibinfo{volume}{58},
  \bibinfo{pages}{7--19}.
\bibitem[{Clayton et~al.(2003)Clayton, Bussey and Dickinson}]{clayton2003can}
\bibinfo{author}{Clayton, N.S.}, \bibinfo{author}{Bussey, T.J.},
  \bibinfo{author}{Dickinson, A.}, \bibinfo{year}{2003}.
\newblock \bibinfo{title}{Can animals recall the past and plan for the future?}
\newblock \bibinfo{journal}{Nature Reviews Neuroscience} \bibinfo{volume}{4},
  \bibinfo{pages}{685--691}.
\bibitem[{Clayton and Dickinson(1998)}]{clayton1998episodic}
\bibinfo{author}{Clayton, N.S.}, \bibinfo{author}{Dickinson, A.},
  \bibinfo{year}{1998}.
\newblock \bibinfo{title}{Episodic-like memory during cache recovery by scrub
  jays}.
\newblock \bibinfo{journal}{Nature} \bibinfo{volume}{395},
  \bibinfo{pages}{272--274}.
\bibitem[{Conard(2003)}]{conard2003palaeolithic}
\bibinfo{author}{Conard, N.J.}, \bibinfo{year}{2003}.
\newblock \bibinfo{title}{Palaeolithic ivory sculptures from southwestern
  germany and the origins of figurative art}.
\newblock \bibinfo{journal}{Nature} \bibinfo{volume}{426},
  \bibinfo{pages}{830--832}.
\bibitem[{Coolidge and Wynn(2005)}]{coolidge2005working}
\bibinfo{author}{Coolidge, F.L.}, \bibinfo{author}{Wynn, T.},
  \bibinfo{year}{2005}.
\newblock \bibinfo{title}{Working memory, its executive functions, and the
  emergence of modern thinking}.
\newblock \bibinfo{journal}{Cambridge archaeological journal}
  \bibinfo{volume}{15}, \bibinfo{pages}{5--26}.
\bibitem[{Crandall et~al.(2018)Crandall, Oudah, Tennom, Ishowo-Oloko, Abdallah,
  Bonnefon, Cebrian, Shariff, Goodrich and Rahwan}]{crandall2018cooperating}
\bibinfo{author}{Crandall, J.W.}, \bibinfo{author}{Oudah, M.},
  \bibinfo{author}{Tennom}, \bibinfo{author}{Ishowo-Oloko, F.},
  \bibinfo{author}{Abdallah, S.}, \bibinfo{author}{Bonnefon, J.F.},
  \bibinfo{author}{Cebrian, M.}, \bibinfo{author}{Shariff, A.},
  \bibinfo{author}{Goodrich, M.A.}, \bibinfo{author}{Rahwan, I.},
  \bibinfo{year}{2018}.
\newblock \bibinfo{title}{Cooperating with machines}.
\newblock \bibinfo{journal}{Nature Communications} \bibinfo{volume}{9},
  \bibinfo{pages}{233}.
\bibitem[{Croft(2000)}]{croft2000explaining}
\bibinfo{author}{Croft, W.}, \bibinfo{year}{2000}.
\newblock \bibinfo{title}{Explaining language change: An evolutionary
  approach}.
\newblock \bibinfo{publisher}{Pearson Education}.
\bibitem[{Croft(2008)}]{croft2008evolutionary}
\bibinfo{author}{Croft, W.}, \bibinfo{year}{2008}.
\newblock \bibinfo{title}{Evolutionary linguistics}.
\newblock \bibinfo{journal}{Annual Review of Anthropology}
  \bibinfo{volume}{37}, \bibinfo{pages}{219--234}.
\bibitem[{Daly(1977)}]{daly1977steady}
\bibinfo{author}{Daly, H.}, \bibinfo{year}{1977}.
\newblock \bibinfo{title}{Steady state economy}.
\newblock \bibinfo{journal}{San Francisco} \bibinfo{volume}{545}.
\bibitem[{Davidson(2006)}]{davidson2006regulatory}
\bibinfo{author}{Davidson, E.H.}, \bibinfo{year}{2006}.
\newblock \bibinfo{title}{The regulatory genome}. volume \bibinfo{volume}{785}.
\newblock \bibinfo{publisher}{Elsevier}.
\bibitem[{Davies(1999)}]{davies1999fifth}
\bibinfo{author}{Davies, P.C.W.}, \bibinfo{year}{1999}.
\newblock \bibinfo{title}{The fifth miracle: The search for the origin and
  meaning of life}.
\newblock \bibinfo{publisher}{Simon and Schuster}.
\bibitem[{Davies(2010)}]{davies2010eerie}
\bibinfo{author}{Davies, P.C.W.}, \bibinfo{year}{2010}.
\newblock \bibinfo{title}{The eerie silence}.
\newblock \bibinfo{journal}{Physics World} \bibinfo{volume}{23},
  \bibinfo{pages}{28}.
\bibitem[{Dawkins(1976)}]{dawkins1976selfish}
\bibinfo{author}{Dawkins, R.}, \bibinfo{year}{1976}.
\newblock \bibinfo{title}{The selfish gene}.
\bibitem[{De~Duve(2005)}]{de2005singularities}
\bibinfo{author}{De~Duve, C.}, \bibinfo{year}{2005}.
\newblock \bibinfo{title}{Singularities: landmarks on the pathways of life}.
\newblock \bibinfo{publisher}{Cambridge University Press}.
\bibitem[{Deacon(1998)}]{deacon1998symbolic}
\bibinfo{author}{Deacon, T.W.}, \bibinfo{year}{1998}.
\newblock \bibinfo{title}{The symbolic species: The co-evolution of language
  and the brain}.
\newblock \bibinfo{publisher}{WW Norton \& Company}.
\bibitem[{Deacon(2011)}]{deacon2011incomplete}
\bibinfo{author}{Deacon, T.W.}, \bibinfo{year}{2011}.
\newblock \bibinfo{title}{Incomplete nature: How mind emerged from matter}.
\newblock \bibinfo{publisher}{WW Norton \& Company}.
\bibitem[{Donahue et~al.(2020)Donahue, Hauser, Nowak and
  Hilbe}]{donahue2020evolving}
\bibinfo{author}{Donahue, K.}, \bibinfo{author}{Hauser, O.P.},
  \bibinfo{author}{Nowak, M.A.}, \bibinfo{author}{Hilbe, C.},
  \bibinfo{year}{2020}.
\newblock \bibinfo{title}{Evolving cooperation in multichannel games}.
\newblock \bibinfo{journal}{Nature Communications} \bibinfo{volume}{11},
  \bibinfo{pages}{3885}.
\bibitem[{Donald(1993)}]{donald1993origins}
\bibinfo{author}{Donald, M.}, \bibinfo{year}{1993}.
\newblock \bibinfo{title}{Origins of the modern mind: Three stages in the
  evolution of culture and cognition}.
\newblock \bibinfo{publisher}{Harvard university press}.
\bibitem[{Doolittle and Bapteste(2007)}]{doolittle2007pattern}
\bibinfo{author}{Doolittle, W.F.}, \bibinfo{author}{Bapteste, E.},
  \bibinfo{year}{2007}.
\newblock \bibinfo{title}{Pattern pluralism and the tree of life hypothesis}.
\newblock \bibinfo{journal}{Proceedings of the National Academy of Sciences}
  \bibinfo{volume}{104}, \bibinfo{pages}{2043--2049}.
\bibitem[{Dunbar(1992)}]{dunbar1992neocortex}
\bibinfo{author}{Dunbar, R.I.}, \bibinfo{year}{1992}.
\newblock \bibinfo{title}{Neocortex size as a constraint on group size in
  primates}.
\newblock \bibinfo{journal}{Journal of human evolution} \bibinfo{volume}{22},
  \bibinfo{pages}{469--493}.
\bibitem[{Dyer(2002)}]{dyer2002biology}
\bibinfo{author}{Dyer, F.C.}, \bibinfo{year}{2002}.
\newblock \bibinfo{title}{The biology of the dance language}.
\newblock \bibinfo{journal}{Annual Review of Entomology} \bibinfo{volume}{47},
  \bibinfo{pages}{917--949}.
\bibitem[{Dyson(1985)}]{dyson1985origins}
\bibinfo{author}{Dyson, F.}, \bibinfo{year}{1985}.
\newblock \bibinfo{title}{Origins of life}.
\newblock \bibinfo{publisher}{Cambridge University Press}.
\bibitem[{Eigen(1971)}]{eigen1971selforganization}
\bibinfo{author}{Eigen, M.}, \bibinfo{year}{1971}.
\newblock \bibinfo{title}{Selforganization of matter and the evolution of
  biological macromolecules}.
\newblock \bibinfo{journal}{Naturwissenschaften} \bibinfo{volume}{58},
  \bibinfo{pages}{465--523}.
\bibitem[{Eigen et~al.(1989)Eigen, McCaskill and Schuster}]{eigen1989molecular}
\bibinfo{author}{Eigen, M.}, \bibinfo{author}{McCaskill, J.},
  \bibinfo{author}{Schuster, P.}, \bibinfo{year}{1989}.
\newblock \bibinfo{title}{The molecular quasi-species}.
\newblock \bibinfo{journal}{Advances in Chemical Physics} \bibinfo{volume}{75},
  \bibinfo{pages}{149--263}.
\bibitem[{Ellis(2012)}]{ellis2012top}
\bibinfo{author}{Ellis, G.F.}, \bibinfo{year}{2012}.
\newblock \bibinfo{title}{Top-down causation and emergence: some comments on
  mechanisms}.
\newblock \bibinfo{journal}{Interface Focus} \bibinfo{volume}{2},
  \bibinfo{pages}{126--140}.
\bibitem[{Engesser et~al.(2024)Engesser, Ridley, Watson, Kita and
  Townsend}]{engesser2024seeds}
\bibinfo{author}{Engesser, S.}, \bibinfo{author}{Ridley, A.R.},
  \bibinfo{author}{Watson, S.K.}, \bibinfo{author}{Kita, S.},
  \bibinfo{author}{Townsend, S.W.}, \bibinfo{year}{2024}.
\newblock \bibinfo{title}{Seeds of language-like generativity in bird call
  combinations}.
\newblock \bibinfo{journal}{Proceedings of the Royal Society B: Biological
  Sciences} \bibinfo{volume}{291}, \bibinfo{pages}{20240922}.
\bibitem[{Evans et~al.(2026)Evans, Bratton and Ag{\"u}era~y
  Arcas}]{evans2026agentic}
\bibinfo{author}{Evans, J.}, \bibinfo{author}{Bratton, B.},
  \bibinfo{author}{Ag{\"u}era~y Arcas, B.}, \bibinfo{year}{2026}.
\newblock \bibinfo{title}{Agentic {AI} and the next intelligence explosion}.
\bibitem[{Falt{\`y}nek et~al.(2019)Falt{\`y}nek, Matlach and
  Lackov{\'a}}]{faltynek2019bases}
\bibinfo{author}{Falt{\`y}nek, D.}, \bibinfo{author}{Matlach, V.},
  \bibinfo{author}{Lackov{\'a}, L.}, \bibinfo{year}{2019}.
\newblock \bibinfo{title}{Bases are not letters: On the analogy between the
  genetic code and natural language by sequence analysis}.
\newblock \bibinfo{journal}{Biosemiotics} \bibinfo{volume}{12},
  \bibinfo{pages}{289--304}.
\bibitem[{Fang et~al.(2002)Fang, Kimbrough, Pace, Valluri and
  Zheng}]{fang2002adaptive}
\bibinfo{author}{Fang, C.}, \bibinfo{author}{Kimbrough, S.O.},
  \bibinfo{author}{Pace, S.}, \bibinfo{author}{Valluri, A.},
  \bibinfo{author}{Zheng, Z.}, \bibinfo{year}{2002}.
\newblock \bibinfo{title}{On adaptive emergence of trust behavior in the game
  of stag hunt}.
\newblock \bibinfo{journal}{Group Decision and Negotiation}
  \bibinfo{volume}{11}, \bibinfo{pages}{449--467}.
\bibitem[{{Ferrer-i-Cancho}(2005)}]{cancho2005variation}
\bibinfo{author}{{Ferrer-i-Cancho}, R.}, \bibinfo{year}{2005}.
\newblock \bibinfo{title}{The variation of {Z}ipf's law in human language}.
\newblock \bibinfo{journal}{The European Physical Journal B-Condensed Matter
  and Complex Systems} \bibinfo{volume}{44}, \bibinfo{pages}{249--257}.
\bibitem[{{Ferrer-i-Cancho}(2025)}]{ferrer2025class}
\bibinfo{author}{{Ferrer-i-Cancho}, R.}, \bibinfo{year}{2025}.
\newblock \bibinfo{title}{On the class of coding optimality of human languages
  and the origins of zipf's law}.
\newblock \bibinfo{journal}{Europhysics Letters} \bibinfo{volume}{151},
  \bibinfo{pages}{62001}.
\bibitem[{{Ferrer-i-Cancho} and D{\'\i}az-Guilera(2007)}]{cancho2007global}
\bibinfo{author}{{Ferrer-i-Cancho}, R.}, \bibinfo{author}{D{\'\i}az-Guilera,
  A.}, \bibinfo{year}{2007}.
\newblock \bibinfo{title}{The global minima of the communicative energy of
  natural communication systems}.
\newblock \bibinfo{journal}{Journal of Statistical Mechanics: Theory and
  Experiment} \bibinfo{volume}{2007}, \bibinfo{pages}{P06009}.
\bibitem[{{Ferrer-i-Cancho} et~al.(2005){Ferrer-i-Cancho}, Riordan and
  Bollob{\'a}s}]{i2005consequences}
\bibinfo{author}{{Ferrer-i-Cancho}, R.}, \bibinfo{author}{Riordan, O.},
  \bibinfo{author}{Bollob{\'a}s, B.}, \bibinfo{year}{2005}.
\newblock \bibinfo{title}{The consequences of {Zipf's} law for syntax and
  symbolic reference}.
\newblock \bibinfo{journal}{Proceedings of the Royal Society B: Biological
  Sciences} \bibinfo{volume}{272}, \bibinfo{pages}{561}.
\bibitem[{{Ferrer-i-Cancho} and Sol{\'e}(2003)}]{cancho2003least}
\bibinfo{author}{{Ferrer-i-Cancho}, R.}, \bibinfo{author}{Sol{\'e}, R.V.},
  \bibinfo{year}{2003}.
\newblock \bibinfo{title}{Least effort and the origins of scaling in human
  language}.
\newblock \bibinfo{journal}{Proceedings of the National Academy of Sciences}
  \bibinfo{volume}{100}, \bibinfo{pages}{788--791}.
\bibitem[{Fisher(2019)}]{fisher2019perils}
\bibinfo{author}{Fisher, R.}, \bibinfo{year}{2019}.
\newblock \bibinfo{title}{The perils of short-termism: Civilisation’s
  greatest threat}.
\newblock \bibinfo{howpublished}{BBC Future}.
\newblock \URLprefix
  \url{https://www.bbc.com/future/article/20190109-the-perils-of-short-termism-civilisations-greatest-threat}.
  \bibinfo{note}{[Online; accessed 11-January-2026]}.
\bibitem[{Fisher(2023)}]{fisher2023long}
\bibinfo{author}{Fisher, R.}, \bibinfo{year}{2023}.
\newblock \bibinfo{title}{The long view: Why we need to transform how the world
  sees time}.
\newblock \bibinfo{publisher}{Hachette UK}.
\bibitem[{Fitch(2010)}]{fitch2010evolution}
\bibinfo{author}{Fitch, W.T.}, \bibinfo{year}{2010}.
\newblock \bibinfo{title}{The evolution of language}.
\newblock \bibinfo{publisher}{Cambridge University Press}.
\bibitem[{Flack(2017)}]{flack2017coarse}
\bibinfo{author}{Flack, J.C.}, \bibinfo{year}{2017}.
\newblock \bibinfo{title}{Coarse-graining as a downward causation mechanism}.
\newblock \bibinfo{journal}{Philosophical Transactions of the Royal Society A}
  \bibinfo{volume}{375}, \bibinfo{pages}{20160338}.
\bibitem[{Fletcher(1993)}]{fletcher1993evolution}
\bibinfo{author}{Fletcher, R.J.}, \bibinfo{year}{1993}.
\newblock \bibinfo{title}{The evolution of human behavior}, in:
  \bibinfo{editor}{Burenhult, G.} (Ed.), \bibinfo{booktitle}{The first humans:
  human origins and history to 10,000 BC}. \bibinfo{publisher}{Harper}, pp.
  \bibinfo{pages}{47--51}.
\bibitem[{Fletcher(1995)}]{fletcher1995limits}
\bibinfo{author}{Fletcher, R.J.}, \bibinfo{year}{1995}.
\newblock \bibinfo{title}{The limits of settlement growth: a theoretical
  outline}.
\newblock \bibinfo{publisher}{Cambridge University Press}.
\bibitem[{Frisch(1993)}]{frisch1993dance}
\bibinfo{author}{Frisch, K.v.}, \bibinfo{year}{1993}.
\newblock \bibinfo{title}{The dance language and orientation of bees}.
\newblock \bibinfo{publisher}{Harvard University Press}.
\bibitem[{Geenen(2021)}]{geenen2021thymus}
\bibinfo{author}{Geenen, V.}, \bibinfo{year}{2021}.
\newblock \bibinfo{title}{The thymus and the science of self}.
\newblock \bibinfo{journal}{Seminars in Immunopathology} \bibinfo{volume}{43},
  \bibinfo{pages}{5--14}.
\bibitem[{Gianni et~al.(2026)Gianni, Kwok, Wan, Goeij, Clifton, Colizzi,
  Attwater and Holliger}]{gianni2026small}
\bibinfo{author}{Gianni, E.}, \bibinfo{author}{Kwok, S.L.Y.},
  \bibinfo{author}{Wan, C.J.K.}, \bibinfo{author}{Goeij, K.},
  \bibinfo{author}{Clifton, B.E.}, \bibinfo{author}{Colizzi, E.S.},
  \bibinfo{author}{Attwater, J.}, \bibinfo{author}{Holliger, P.},
  \bibinfo{year}{2026}.
\newblock \bibinfo{title}{A small polymerase ribozyme that can synthesize
  itself and its complementary strand}.
\newblock \bibinfo{journal}{Science} \bibinfo{volume}{0},
  \bibinfo{pages}{eadt2760}.
\newblock \DOIprefix\doi{10.1126/science.adt2760}.
\bibitem[{Gibbons(2018)}]{gibbons2018complex}
\bibinfo{author}{Gibbons, A.}, \bibinfo{year}{2018}.
\newblock \bibinfo{title}{Complex behavior arose at dawn of humans}.
\newblock \bibinfo{journal}{Science} \bibinfo{volume}{359},
  \bibinfo{pages}{1200--1201}.
\bibitem[{Gilbert(1986)}]{gilbert1986origin}
\bibinfo{author}{Gilbert, W.}, \bibinfo{year}{1986}.
\newblock \bibinfo{title}{Origin of life: The rna world}.
\newblock \bibinfo{journal}{Nature} \bibinfo{volume}{319},
  \bibinfo{pages}{618--618}.
\bibitem[{Gogarten et~al.(2002)Gogarten, Doolittle and
  Lawrence}]{gogarten2002prokaryotic}
\bibinfo{author}{Gogarten, J.P.}, \bibinfo{author}{Doolittle, W.F.},
  \bibinfo{author}{Lawrence, J.G.}, \bibinfo{year}{2002}.
\newblock \bibinfo{title}{Prokaryotic evolution in light of gene transfer}.
\newblock \bibinfo{journal}{Molecular Biology and Evolution}
  \bibinfo{volume}{19}, \bibinfo{pages}{2226--2238}.
\bibitem[{Goody(1986)}]{goody1986logic}
\bibinfo{author}{Goody, J.}, \bibinfo{year}{1986}.
\newblock \bibinfo{title}{The logic of writing and the organization of
  society}.
\newblock \bibinfo{publisher}{Cambridge University Press}.
\bibitem[{Grush(2004)}]{grush2004emulation}
\bibinfo{author}{Grush, R.}, \bibinfo{year}{2004}.
\newblock \bibinfo{title}{The emulation theory of representation: Motor
  control, imagery, and perception}.
\newblock \bibinfo{journal}{Behavioral and Brain Sciences}
  \bibinfo{volume}{27}, \bibinfo{pages}{377--396}.
\bibitem[{Hanson(1998)}]{hanson1998great}
\bibinfo{author}{Hanson, R.}, \bibinfo{year}{1998}.
\newblock \bibinfo{title}{{The Great Filter --- Are we almost past it?}}
\newblock \URLprefix \url{https://mason.gmu.edu/~rhanson/greatfilter.html}.
  \bibinfo{note}{[Online; accessed 15-Sept-2025]}.
\bibitem[{Harari(2023)}]{harari2023yuval}
\bibinfo{author}{Harari, Y.N.}, \bibinfo{year}{2023}.
\newblock \bibinfo{title}{Yuval noah harari argues that ai has hacked the
  operating system of human civilisation}.
\newblock \bibinfo{journal}{The Economist} \bibinfo{volume}{28}.
\bibitem[{Harr{\'e} et~al.(2024)Harr{\'e}, Drysdale and
  Ruiz-Serra}]{harre2024ai}
\bibinfo{author}{Harr{\'e}, M.S.}, \bibinfo{author}{Drysdale, C.},
  \bibinfo{author}{Ruiz-Serra, J.}, \bibinfo{year}{2024}.
\newblock \bibinfo{title}{An {AI} theory of mind will enhance our collective
  intelligence}.
\newblock \bibinfo{journal}{arXiv preprint arXiv:2411.09168} .
\bibitem[{Harr{\'e} and Prokopenko(2016)}]{harre2016social}
\bibinfo{author}{Harr{\'e}, M.S.}, \bibinfo{author}{Prokopenko, M.},
  \bibinfo{year}{2016}.
\newblock \bibinfo{title}{The social brain: scale-invariant layering of
  {Erd{\H{o}}s--R{\'e}nyi} networks in small-scale human societies}.
\newblock \bibinfo{journal}{Journal of the Royal Society Interface}
  \bibinfo{volume}{13}, \bibinfo{pages}{20160044}.
\bibitem[{Hauert et~al.(2006)Hauert, Holmes and
  Doebeli}]{hauert2006evolutionary}
\bibinfo{author}{Hauert, C.}, \bibinfo{author}{Holmes, M.},
  \bibinfo{author}{Doebeli, M.}, \bibinfo{year}{2006}.
\newblock \bibinfo{title}{Evolutionary games and population dynamics:
  maintenance of cooperation in public goods games}.
\newblock \bibinfo{journal}{Proceedings of the Royal Society B: Biological
  Sciences} \bibinfo{volume}{273}, \bibinfo{pages}{2565--2571}.
\bibitem[{Hauert et~al.(2008)Hauert, Wakano and Doebeli}]{hauert2008ecological}
\bibinfo{author}{Hauert, C.}, \bibinfo{author}{Wakano, J.Y.},
  \bibinfo{author}{Doebeli, M.}, \bibinfo{year}{2008}.
\newblock \bibinfo{title}{Ecological public goods games: cooperation and
  bifurcation}.
\newblock \bibinfo{journal}{Theoretical Population Biology}
  \bibinfo{volume}{73}, \bibinfo{pages}{257--263}.
\bibitem[{Hauser et~al.(2002)Hauser, Chomsky and Fitch}]{hauser2002faculty}
\bibinfo{author}{Hauser, M.D.}, \bibinfo{author}{Chomsky, N.},
  \bibinfo{author}{Fitch, W.T.}, \bibinfo{year}{2002}.
\newblock \bibinfo{title}{The faculty of language: what is it, who has it, and
  how did it evolve?}
\newblock \bibinfo{journal}{Science} \bibinfo{volume}{298},
  \bibinfo{pages}{1569--1579}.
\bibitem[{Heinsalu et~al.(2026)Heinsalu, Patriarca and
  Sanchez}]{heinsalu2025language}
\bibinfo{author}{Heinsalu, E.}, \bibinfo{author}{Patriarca, M.},
  \bibinfo{author}{Sanchez, D.}, \bibinfo{year}{2026}.
\newblock \bibinfo{title}{Language dynamics: Complex systems approaches}.
\newblock \bibinfo{journal}{International Encyclopedia of Language and
  Linguistics} \bibinfo{volume}{14}, \bibinfo{pages}{241--247}.
\bibitem[{Henshilwood et~al.(2001)Henshilwood, d'Errico, Marean, Milo and
  Yates}]{henshilwood2001early}
\bibinfo{author}{Henshilwood, C.S.}, \bibinfo{author}{d'Errico, F.},
  \bibinfo{author}{Marean, C.W.}, \bibinfo{author}{Milo, R.G.},
  \bibinfo{author}{Yates, R.}, \bibinfo{year}{2001}.
\newblock \bibinfo{title}{An early bone tool industry from the middle stone age
  at blombos cave, south africa: implications for the origins of modern human
  behaviour, symbolism and language}.
\newblock \bibinfo{journal}{Journal of Human Evolution} \bibinfo{volume}{41},
  \bibinfo{pages}{631--678}.
\bibitem[{Henshilwood and Dubreuil(2009)}]{henshilwood2009reading}
\bibinfo{author}{Henshilwood, C.S.}, \bibinfo{author}{Dubreuil, B.},
  \bibinfo{year}{2009}.
\newblock \bibinfo{title}{Reading the artifacts: gleaning language skills from
  the middle stone age in southern africa}, in: \bibinfo{editor}{Botha, R.},
  \bibinfo{editor}{Knight, C.} (Eds.), \bibinfo{booktitle}{The Cradle of
  Language}. \bibinfo{publisher}{Oxford University Press}, pp.
  \bibinfo{pages}{41--61}.
\bibitem[{Hern{\'a}ndez-Orozco et~al.(2018)Hern{\'a}ndez-Orozco,
  Hern{\'a}ndez-Quiroz and Zenil}]{hernandez2018undecidability}
\bibinfo{author}{Hern{\'a}ndez-Orozco, S.},
  \bibinfo{author}{Hern{\'a}ndez-Quiroz, F.}, \bibinfo{author}{Zenil, H.},
  \bibinfo{year}{2018}.
\newblock \bibinfo{title}{Undecidability and irreducibility conditions for
  open-ended evolution and emergence}.
\newblock \bibinfo{journal}{Artificial Life} \bibinfo{volume}{24},
  \bibinfo{pages}{56--70}.
\bibitem[{Heyes(2018)}]{heyes2018cognitive}
\bibinfo{author}{Heyes, C.}, \bibinfo{year}{2018}.
\newblock \bibinfo{title}{Cognitive gadgets: The cultural evolution of
  thinking}.
\newblock \bibinfo{publisher}{Harvard University Press}.
\bibitem[{Hoel(2017)}]{hoel2017map}
\bibinfo{author}{Hoel, E.P.}, \bibinfo{year}{2017}.
\newblock \bibinfo{title}{When the map is better than the territory}.
\newblock \bibinfo{journal}{Entropy} \bibinfo{volume}{19},
  \bibinfo{pages}{188}.
\bibitem[{Hoel(2026)}]{hoel2026quantifying}
\bibinfo{author}{Hoel, E.P.}, \bibinfo{year}{2026}.
\newblock \bibinfo{title}{Quantifying emergent complexity}.
\newblock \bibinfo{journal}{Patterns} \bibinfo{volume}{7}.
\bibitem[{Hoffmeyer(1997)}]{hoffmeyer1997signs}
\bibinfo{author}{Hoffmeyer, J.}, \bibinfo{year}{1997}.
\newblock \bibinfo{title}{Signs of Meaning in the Universe}.
\newblock \bibinfo{publisher}{Indiana University Press}.
\bibitem[{Hofstadter(1980)}]{hofstadter_go_1980}
\bibinfo{author}{Hofstadter, D.R.}, \bibinfo{year}{1980}.
\newblock \bibinfo{title}{G\"odel, {{Escher}}, {{Bach}}: An Eternal Golden
  Braid}.
\newblock \bibinfo{publisher}{{Penguin}}, \bibinfo{address}{{Harmondsworth}}.
\bibitem[{Hopfield(1974)}]{hopfield1974kinetic}
\bibinfo{author}{Hopfield, J.J.}, \bibinfo{year}{1974}.
\newblock \bibinfo{title}{Kinetic proofreading: a new mechanism for reducing
  errors in biosynthetic processes requiring high specificity}.
\newblock \bibinfo{journal}{Proceedings of the National Academy of Sciences}
  \bibinfo{volume}{71}, \bibinfo{pages}{4135--4139}.
\bibitem[{Huttegger et~al.(2014)Huttegger, Skyrms, Tarres and
  Wagner}]{huttegger2014some}
\bibinfo{author}{Huttegger, S.}, \bibinfo{author}{Skyrms, B.},
  \bibinfo{author}{Tarres, P.}, \bibinfo{author}{Wagner, E.},
  \bibinfo{year}{2014}.
\newblock \bibinfo{title}{Some dynamics of signaling games}.
\newblock \bibinfo{journal}{Proceedings of the National Academy of Sciences}
  \bibinfo{volume}{111}, \bibinfo{pages}{10873--10880}.
\bibitem[{Igamberdiev(2021)}]{igamberdiev2021drawbridge}
\bibinfo{author}{Igamberdiev, A.U.}, \bibinfo{year}{2021}.
\newblock \bibinfo{title}{The drawbridge of nature: Evolutionary
  complexification as a generation and novel interpretation of coding systems}.
\newblock \bibinfo{journal}{Biosystems} \bibinfo{volume}{207},
  \bibinfo{pages}{104454}.
\bibitem[{Jablonka(1994)}]{jablonka1994inheritance}
\bibinfo{author}{Jablonka, E.}, \bibinfo{year}{1994}.
\newblock \bibinfo{title}{Inheritance systems and the evolution of new levels
  of individuality}.
\newblock \bibinfo{journal}{Journal of Theoretical Biology}
  \bibinfo{volume}{170}, \bibinfo{pages}{301--309}.
\bibitem[{Jackendoff(1999)}]{jackendoff1999possible}
\bibinfo{author}{Jackendoff, R.}, \bibinfo{year}{1999}.
\newblock \bibinfo{title}{Possible stages in the evolution of the language
  capacity}.
\newblock \bibinfo{journal}{Trends in Cognitive Sciences} \bibinfo{volume}{3},
  \bibinfo{pages}{272--279}.
\bibitem[{Jakobson(1971)}]{jakobson1971selected}
\bibinfo{author}{Jakobson, R.}, \bibinfo{year}{1971}.
\newblock \bibinfo{title}{Selected writings [of] Roman Jakobson: Word and
  language}.
\newblock \bibinfo{publisher}{Mouton}.
\bibitem[{Joyce(2002)}]{joyce2002antiquity}
\bibinfo{author}{Joyce, G.F.}, \bibinfo{year}{2002}.
\newblock \bibinfo{title}{The antiquity of rna-based evolution}.
\newblock \bibinfo{journal}{Nature} \bibinfo{volume}{418},
  \bibinfo{pages}{214--221}.
\bibitem[{Kallis et~al.(2025)Kallis, Hickel, O’Neill, Jackson, Victor,
  Raworth, Schor, Steinberger and {\"U}rge-Vorsatz}]{kallis2025post}
\bibinfo{author}{Kallis, G.}, \bibinfo{author}{Hickel, J.},
  \bibinfo{author}{O’Neill, D.W.}, \bibinfo{author}{Jackson, T.},
  \bibinfo{author}{Victor, P.A.}, \bibinfo{author}{Raworth, K.},
  \bibinfo{author}{Schor, J.B.}, \bibinfo{author}{Steinberger, J.K.},
  \bibinfo{author}{{\"U}rge-Vorsatz, D.}, \bibinfo{year}{2025}.
\newblock \bibinfo{title}{Post-growth: the science of wellbeing within
  planetary boundaries}.
\newblock \bibinfo{journal}{The Lancet Planetary Health} \bibinfo{volume}{9},
  \bibinfo{pages}{e62--e78}.
\bibitem[{Kauffman(1986)}]{kauffman1986autocatalytic}
\bibinfo{author}{Kauffman, S.A.}, \bibinfo{year}{1986}.
\newblock \bibinfo{title}{Autocatalytic sets of proteins}.
\newblock \bibinfo{journal}{Journal of Theoretical Biology}
  \bibinfo{volume}{119}, \bibinfo{pages}{1--24}.
\bibitem[{Kauffman(1993)}]{kauffman1993origins}
\bibinfo{author}{Kauffman, S.A.}, \bibinfo{year}{1993}.
\newblock \bibinfo{title}{The origins of order: Self-organization and selection
  in evolution}.
\newblock \bibinfo{publisher}{Oxford University Press}.
\bibitem[{Key{\ss}er and Lenzen(2021)}]{keysser20211}
\bibinfo{author}{Key{\ss}er, L.T.}, \bibinfo{author}{Lenzen, M.},
  \bibinfo{year}{2021}.
\newblock \bibinfo{title}{{1.5 C degrowth scenarios suggest the need for new
  mitigation pathways}}.
\newblock \bibinfo{journal}{Nature Communications} \bibinfo{volume}{12},
  \bibinfo{pages}{2676}.
\bibitem[{Khersonsky et~al.(2006)Khersonsky, Roodveldt and
  Tawfik}]{khersonsky2006enzyme}
\bibinfo{author}{Khersonsky, O.}, \bibinfo{author}{Roodveldt, C.},
  \bibinfo{author}{Tawfik, D.S.}, \bibinfo{year}{2006}.
\newblock \bibinfo{title}{Enzyme promiscuity: evolutionary and mechanistic
  aspects}.
\newblock \bibinfo{journal}{Current Opinion in Chemical Biology}
  \bibinfo{volume}{10}, \bibinfo{pages}{498--508}.
\bibitem[{Kim et~al.(2026)Kim, Lai, Scherrer, Evans et~al.}]{kim2026reasoning}
\bibinfo{author}{Kim, J.}, \bibinfo{author}{Lai, S.},
  \bibinfo{author}{Scherrer, N.}, \bibinfo{author}{Evans, J.}, et~al.,
  \bibinfo{year}{2026}.
\newblock \bibinfo{title}{Reasoning models generate societies of thought}.
\newblock \bibinfo{journal}{arXiv preprint arXiv:2601.10825} .
\bibitem[{Kun(2021)}]{kun2021major}
\bibinfo{author}{Kun, {\'A}.}, \bibinfo{year}{2021}.
\newblock \bibinfo{title}{The major evolutionary transitions and codes of
  life}.
\newblock \bibinfo{journal}{Biosystems} \bibinfo{volume}{210},
  \bibinfo{pages}{104548}.
\bibitem[{Lee et~al.(2023)Lee, Calvin, Dasgupta, Krinner, Mukherji, Thorne,
  Trisos, Romero, Aldunce, Barret et~al.}]{lee2023ipcc}
\bibinfo{author}{Lee, H.}, \bibinfo{author}{Calvin, K.},
  \bibinfo{author}{Dasgupta, D.}, \bibinfo{author}{Krinner, G.},
  \bibinfo{author}{Mukherji, A.}, \bibinfo{author}{Thorne, P.},
  \bibinfo{author}{Trisos, C.}, \bibinfo{author}{Romero, J.},
  \bibinfo{author}{Aldunce, P.}, \bibinfo{author}{Barret, K.}, et~al.,
  \bibinfo{year}{2023}.
\newblock \bibinfo{title}{IPCC, 2023: Climate change 2023: Synthesis report,
  summary for policymakers. Contribution of working groups I, II and III to the
  Sixth Assessment Report of the Intergovernmental Panel on Climate Change
  [Core writing team, H. Lee and J. Romero (eds.)]. IPCC, Geneva, Switzerland.}
\newblock \bibinfo{type}{Technical Report}. Intergovernmental Panel on Climate
  Change (IPCC).
\bibitem[{Lewis(2008)}]{lewis2008convention}
\bibinfo{author}{Lewis, D.}, \bibinfo{year}{2008}.
\newblock \bibinfo{title}{Convention: A philosophical study}.
\newblock \bibinfo{publisher}{John Wiley \& Sons}.
\bibitem[{Lingam and Loeb(2019)}]{lingam2019role}
\bibinfo{author}{Lingam, M.}, \bibinfo{author}{Loeb, A.}, \bibinfo{year}{2019}.
\newblock \bibinfo{title}{Role of stellar physics in regulating the critical
  steps for life}.
\newblock \bibinfo{journal}{International Journal of Astrobiology}
  \bibinfo{volume}{18}, \bibinfo{pages}{527--546}.
\bibitem[{Lingam and Loeb(2021)}]{lingam2021life}
\bibinfo{author}{Lingam, M.}, \bibinfo{author}{Loeb, A.}, \bibinfo{year}{2021}.
\newblock \bibinfo{title}{Life in the cosmos: From biosignatures to
  technosignatures}.
\newblock \bibinfo{publisher}{Harvard University Press}.
\bibitem[{Malik et~al.(2024)Malik, Lenzen, Li, Mora, Carter, Giljum, Lutter and
  G{\'o}mez-Paredes}]{malik2024polarizing}
\bibinfo{author}{Malik, A.}, \bibinfo{author}{Lenzen, M.}, \bibinfo{author}{Li,
  M.}, \bibinfo{author}{Mora, C.}, \bibinfo{author}{Carter, S.},
  \bibinfo{author}{Giljum, S.}, \bibinfo{author}{Lutter, S.},
  \bibinfo{author}{G{\'o}mez-Paredes, J.}, \bibinfo{year}{2024}.
\newblock \bibinfo{title}{{Polarizing and equalizing trends in international
  trade and Sustainable Development Goals}}.
\newblock \bibinfo{journal}{Nature Sustainability} \bibinfo{volume}{7},
  \bibinfo{pages}{1359--1370}.
\bibitem[{Mandelbrot(1953)}]{mandelbrot1953informational}
\bibinfo{author}{Mandelbrot, B.}, \bibinfo{year}{1953}.
\newblock \bibinfo{title}{An informational theory of the statistical structure
  of language}.
\newblock \bibinfo{journal}{Communication theory} \bibinfo{volume}{84},
  \bibinfo{pages}{486--502}.
\bibitem[{Manheim(2020)}]{manheim2020fragile}
\bibinfo{author}{Manheim, D.}, \bibinfo{year}{2020}.
\newblock \bibinfo{title}{The fragile world hypothesis: {Complexity},
  fragility, and systemic existential risk}.
\newblock \bibinfo{journal}{Futures} \bibinfo{volume}{122},
  \bibinfo{pages}{102570}.
\bibitem[{Marion(2012)}]{marion2012turing}
\bibinfo{author}{Marion, J.Y.}, \bibinfo{year}{2012}.
\newblock \bibinfo{title}{From turing machines to computer viruses}.
\newblock \bibinfo{journal}{Philosophical Transactions of the Royal Society A:
  Mathematical, Physical and Engineering Sciences} \bibinfo{volume}{370},
  \bibinfo{pages}{3319--3339}.
\bibitem[{Markose(2004)}]{markose2004novelty}
\bibinfo{author}{Markose, S.M.}, \bibinfo{year}{2004}.
\newblock \bibinfo{title}{Novelty in complex adaptive systems (cas) dynamics: a
  computational theory of actor innovation}.
\newblock \bibinfo{journal}{Physica A: Statistical Mechanics and Its
  Applications} \bibinfo{volume}{344}, \bibinfo{pages}{41--49}.
\bibitem[{Markose(2017)}]{markose_complex_2017}
\bibinfo{author}{Markose, S.M.}, \bibinfo{year}{2017}.
\newblock \bibinfo{title}{Complex type 4 structure changing dynamics of digital
  agents: {{Nash}} equilibria of a game with arms race in innovations}.
\newblock \bibinfo{journal}{Journal of Dynamics and Games} \bibinfo{volume}{4},
  \bibinfo{pages}{255--284}.
\bibitem[{Markose(2021)}]{markose2021genomic}
\bibinfo{author}{Markose, S.M.}, \bibinfo{year}{2021}.
\newblock \bibinfo{title}{Genomic intelligence as {\"u}ber bio-cybersecurity:
  The g{\"o}del sentence in immuno-cognitive systems}.
\newblock \bibinfo{journal}{Entropy} \bibinfo{volume}{23},
  \bibinfo{pages}{405}.
\bibitem[{Massey and Mishra(2018)}]{massey2018origin}
\bibinfo{author}{Massey, S.E.}, \bibinfo{author}{Mishra, B.},
  \bibinfo{year}{2018}.
\newblock \bibinfo{title}{Origin of biomolecular games: deception and molecular
  evolution}.
\newblock \bibinfo{journal}{Journal of The Royal Society Interface}
  \bibinfo{volume}{15}.
\bibitem[{Mills et~al.(2025)Mills, Macalady, Frank and
  Wright}]{mills2025reassessment}
\bibinfo{author}{Mills, D.B.}, \bibinfo{author}{Macalady, J.L.},
  \bibinfo{author}{Frank, A.}, \bibinfo{author}{Wright, J.T.},
  \bibinfo{year}{2025}.
\newblock \bibinfo{title}{A reassessment of the ``hard-steps'' model for the
  evolution of intelligent life}.
\newblock \bibinfo{journal}{Science Advances} \bibinfo{volume}{11},
  \bibinfo{pages}{eads5698}.
\bibitem[{Mufwene(2008)}]{mufwene2008language}
\bibinfo{author}{Mufwene, S.S.}, \bibinfo{year}{2008}.
\newblock \bibinfo{title}{Language evolution: Contact, competition and change}.
\newblock \bibinfo{publisher}{Bloomsbury Publishing}.
\bibitem[{Myers(1904)}]{myers1904ancient}
\bibinfo{author}{Myers, P.V.N.}, \bibinfo{year}{1904}.
\newblock \bibinfo{title}{Ancient history}.
\newblock \bibinfo{publisher}{Ginn}.
\bibitem[{{National Human Genome Research Institute}(2026)}]{NHGRI-gen-code}
\bibinfo{author}{{National Human Genome Research Institute}},
  \bibinfo{year}{2026}.
\newblock \bibinfo{title}{Genetic code}.
\newblock \URLprefix
  \url{https://www.genome.gov/genetics-glossary/Genetic-Code}.
  \bibinfo{note}{[Online; accessed 5-Aug-2026]}.
\bibitem[{Obst et~al.(2011)Obst, Polani and Prokopenko}]{obst2011origins}
\bibinfo{author}{Obst, O.}, \bibinfo{author}{Polani, D.},
  \bibinfo{author}{Prokopenko, M.}, \bibinfo{year}{2011}.
\newblock \bibinfo{title}{Origins of scaling in genetic code}, in:
  \bibinfo{booktitle}{Advances in Artificial Life. Darwin Meets von Neumann.
  Proceedings, 10th European Conference, ECAL 2009, Budapest, Hungary,
  September 13-16, 2009}. \bibinfo{publisher}{Springer}, pp.
  \bibinfo{pages}{85--93}.
\bibitem[{Ong(2003)}]{ong2003orality}
\bibinfo{author}{Ong, W.J.}, \bibinfo{year}{2003}.
\newblock \bibinfo{title}{Orality and literacy}.
\newblock \bibinfo{publisher}{Routledge}.
\bibitem[{Pattee(2001)}]{pattee2001physics}
\bibinfo{author}{Pattee, H.H.}, \bibinfo{year}{2001}.
\newblock \bibinfo{title}{The physics of symbols: bridging the epistemic cut}.
\newblock \bibinfo{journal}{Biosystems} \bibinfo{volume}{60},
  \bibinfo{pages}{5--21}.
\bibitem[{Pattee(2012)}]{pattee2012does}
\bibinfo{author}{Pattee, H.H.}, \bibinfo{year}{2012}.
\newblock \bibinfo{title}{How does a molecule become a message?}, in:
  \bibinfo{booktitle}{Laws, Language and Life: Howard Pattee’s classic papers
  on the physics of symbols with contemporary commentary}.
  \bibinfo{publisher}{Springer}, pp. \bibinfo{pages}{55--67}.
\bibitem[{Penrose and Penrose(1958)}]{penrose1958impossible}
\bibinfo{author}{Penrose, L.S.}, \bibinfo{author}{Penrose, R.},
  \bibinfo{year}{1958}.
\newblock \bibinfo{title}{Impossible objects: a special type of visual
  illusion.}
\newblock \bibinfo{journal}{British Journal of Psychology}
  \bibinfo{volume}{49}, \bibinfo{pages}{31--33}.
\bibitem[{Pettitt(2010)}]{pettitt2010palaeolithic}
\bibinfo{author}{Pettitt, P.}, \bibinfo{year}{2010}.
\newblock \bibinfo{title}{The Palaeolithic origins of human burial}.
\newblock \bibinfo{publisher}{Routledge}.
\bibitem[{Phaniraj et~al.(2026)Phaniraj, Wierucka and
  Burkart}]{phaniraj2026dynamic}
\bibinfo{author}{Phaniraj, N.}, \bibinfo{author}{Wierucka, K.},
  \bibinfo{author}{Burkart, J.M.}, \bibinfo{year}{2026}.
\newblock \bibinfo{title}{Dynamic vocal learning in adult marmoset monkeys}.
\newblock \bibinfo{journal}{Proceedings of the Royal Society B: Biological
  Sciences} \bibinfo{volume}{293}, \bibinfo{pages}{20260743}.
\bibitem[{Pinotsis and Miller(2026)}]{a2026ephaptic}
\bibinfo{author}{Pinotsis, D.A.}, \bibinfo{author}{Miller, E.K.},
  \bibinfo{year}{2026}.
\newblock \bibinfo{title}{Ephaptic coupling can explain variability in neural
  activity}.
\newblock \bibinfo{journal}{Cerebral Cortex} \bibinfo{volume}{36},
  \bibinfo{pages}{bhag098}.
\bibitem[{Post(1944)}]{post_recursively_1944}
\bibinfo{author}{Post, E.L.}, \bibinfo{year}{1944}.
\newblock \bibinfo{title}{Recursively enumerable sets of positive integers and
  their decision problems}.
\newblock \bibinfo{journal}{Bulletin of the American Mathematical Society}
  \bibinfo{volume}{50}, \bibinfo{pages}{284--316}.
\bibitem[{Prokopenko et~al.(2010)Prokopenko, Ay, Obst and
  Polani}]{prokopenko2010phase}
\bibinfo{author}{Prokopenko, M.}, \bibinfo{author}{Ay, N.},
  \bibinfo{author}{Obst, O.}, \bibinfo{author}{Polani, D.},
  \bibinfo{year}{2010}.
\newblock \bibinfo{title}{Phase transitions in least-effort communications}.
\newblock \bibinfo{journal}{Journal of Statistical Mechanics: Theory and
  Experiment} \bibinfo{volume}{2010}, \bibinfo{pages}{P11025}.
\bibitem[{Prokopenko et~al.(2025)Prokopenko, Davies, Harr{\'e}, Heisler,
  Kuncic, Lewis, Livson, Lizier and Rosas}]{prokopenko2025biological}
\bibinfo{author}{Prokopenko, M.}, \bibinfo{author}{Davies, P.C.},
  \bibinfo{author}{Harr{\'e}, M.}, \bibinfo{author}{Heisler, M.G.},
  \bibinfo{author}{Kuncic, Z.}, \bibinfo{author}{Lewis, G.F.},
  \bibinfo{author}{Livson, O.}, \bibinfo{author}{Lizier, J.T.},
  \bibinfo{author}{Rosas, F.E.}, \bibinfo{year}{2025}.
\newblock \bibinfo{title}{Biological arrow of time: emergence of tangled
  information hierarchies and self-modelling dynamics}.
\newblock \bibinfo{journal}{Journal of Physics: Complexity}
  \bibinfo{volume}{6}, \bibinfo{pages}{015006}.
\bibitem[{Prokopenko et~al.(2019)Prokopenko, Harr\'e, Lizier, Boschetti, Peppas
  and Kauffman}]{prokopenko_self-referential_2019}
\bibinfo{author}{Prokopenko, M.}, \bibinfo{author}{Harr\'e, M.},
  \bibinfo{author}{Lizier, J.}, \bibinfo{author}{Boschetti, F.},
  \bibinfo{author}{Peppas, P.}, \bibinfo{author}{Kauffman, S.},
  \bibinfo{year}{2019}.
\newblock \bibinfo{title}{Self-referential basis of undecidable dynamics:
  {{From}} the {{Liar}} paradox and the halting problem to the edge of chaos}.
\newblock \bibinfo{journal}{Physics of Life Reviews} \bibinfo{volume}{31},
  \bibinfo{pages}{134--156}.
\bibitem[{Prokopenko et~al.(2009)Prokopenko, Polani and
  Chadwick}]{prokopenko_stigmergic_2009}
\bibinfo{author}{Prokopenko, M.}, \bibinfo{author}{Polani, D.},
  \bibinfo{author}{Chadwick, M.}, \bibinfo{year}{2009}.
\newblock \bibinfo{title}{Stigmergic gene transfer and emergence of universal
  coding}.
\newblock \bibinfo{journal}{HFSP Journal} \bibinfo{volume}{3},
  \bibinfo{pages}{317--327}.
\bibitem[{Rappaport(1999)}]{rappaport1999ritual}
\bibinfo{author}{Rappaport, R.A.}, \bibinfo{year}{1999}.
\newblock \bibinfo{title}{Ritual and Religion in the Making of Humanity}.
  volume \bibinfo{volume}{110}.
\newblock \bibinfo{publisher}{Cambridge University Press}.
\bibitem[{Richerson and Boyd(2000)}]{richerson2000pleistocene}
\bibinfo{author}{Richerson, P.J.}, \bibinfo{author}{Boyd, R.},
  \bibinfo{year}{2000}.
\newblock \bibinfo{title}{The pleistocene and the origins of human culture:
  built for speed}.
\newblock \bibinfo{journal}{Perspectives in Ethology} \bibinfo{volume}{13},
  \bibinfo{pages}{1--45}.
\bibitem[{Rockstr{\"o}m et~al.(2024)Rockstr{\"o}m, Donges, Fetzer, Martin,
  Wang-Erlandsson and Richardson}]{rockstrom2024planetary}
\bibinfo{author}{Rockstr{\"o}m, J.}, \bibinfo{author}{Donges, J.F.},
  \bibinfo{author}{Fetzer, I.}, \bibinfo{author}{Martin, M.A.},
  \bibinfo{author}{Wang-Erlandsson, L.}, \bibinfo{author}{Richardson, K.},
  \bibinfo{year}{2024}.
\newblock \bibinfo{title}{{Planetary Boundaries guide humanity’s future on
  Earth}}.
\newblock \bibinfo{journal}{Nature Reviews Earth \& Environment}
  \bibinfo{volume}{5}, \bibinfo{pages}{773--788}.
\bibitem[{Rodin and Rodin(2008)}]{rodin2008origin}
\bibinfo{author}{Rodin, S.}, \bibinfo{author}{Rodin, A.}, \bibinfo{year}{2008}.
\newblock \bibinfo{title}{On the origin of the genetic code: signatures of its
  primordial complementarity in trnas and aminoacyl-trna synthetases}.
\newblock \bibinfo{journal}{Heredity} \bibinfo{volume}{100},
  \bibinfo{pages}{341--355}.
\bibitem[{Rosen(1991)}]{rosen1991life}
\bibinfo{author}{Rosen, R.}, \bibinfo{year}{1991}.
\newblock \bibinfo{title}{Life itself: A comprehensive inquiry into the nature,
  origin, and fabrication of life}.
\newblock \bibinfo{publisher}{Columbia University Press}.
\bibitem[{Sagan(1963)}]{sagan1963direct}
\bibinfo{author}{Sagan, C.}, \bibinfo{year}{1963}.
\newblock \bibinfo{title}{Direct contact among galactic civilizations by
  relativistic interstellar spaceflight}.
\newblock \bibinfo{journal}{Planetary and Space Science} \bibinfo{volume}{11},
  \bibinfo{pages}{485--498}.
\bibitem[{Salge et~al.(2015)Salge, Ay, Polani and Prokopenko}]{salge2015zipf}
\bibinfo{author}{Salge, C.}, \bibinfo{author}{Ay, N.}, \bibinfo{author}{Polani,
  D.}, \bibinfo{author}{Prokopenko, M.}, \bibinfo{year}{2015}.
\newblock \bibinfo{title}{Zipf's law: balancing signal usage cost and
  communication efficiency}.
\newblock \bibinfo{journal}{PLoS One} \bibinfo{volume}{10},
  \bibinfo{pages}{e0139475}.
\bibitem[{San~Miguel and Toral(2026)}]{san2025individual}
\bibinfo{author}{San~Miguel, M.}, \bibinfo{author}{Toral, R.},
  \bibinfo{year}{2026}.
\newblock \bibinfo{title}{Individual based models of language competition}.
\newblock \bibinfo{journal}{International Encyclopedia of Language and
  Linguistics} \bibinfo{volume}{14}, \bibinfo{pages}{248--257}.
\bibitem[{Sandholm(2005)}]{sandholm2005negative}
\bibinfo{author}{Sandholm, W.H.}, \bibinfo{year}{2005}.
\newblock \bibinfo{title}{Negative externalities and evolutionary
  implementation}.
\newblock \bibinfo{journal}{The Review of Economic Studies}
  \bibinfo{volume}{72}, \bibinfo{pages}{885--915}.
\bibitem[{Sayama(2008)}]{sayama2008construction}
\bibinfo{author}{Sayama, H.}, \bibinfo{year}{2008}.
\newblock \bibinfo{title}{Construction theory, self-replication, and the
  halting problem}.
\newblock \bibinfo{journal}{Complexity} \bibinfo{volume}{13},
  \bibinfo{pages}{16--22}.
\bibitem[{Scherjon et~al.(2015)Scherjon, Bakels, MacDonald and
  Roebroeks}]{scherjon2015burning}
\bibinfo{author}{Scherjon, F.}, \bibinfo{author}{Bakels, C.},
  \bibinfo{author}{MacDonald, K.}, \bibinfo{author}{Roebroeks, W.},
  \bibinfo{year}{2015}.
\newblock \bibinfo{title}{Burning the land: An ethnographic study of off-site
  fire use by current and historically documented foragers and implications for
  the interpretation of past fire practices in the landscape}.
\newblock \bibinfo{journal}{Current Anthropology} \bibinfo{volume}{56},
  \bibinfo{pages}{299--326}.
\bibitem[{Searls(2002)}]{searls2002language}
\bibinfo{author}{Searls, D.B.}, \bibinfo{year}{2002}.
\newblock \bibinfo{title}{The language of genes}.
\newblock \bibinfo{journal}{Nature} \bibinfo{volume}{420},
  \bibinfo{pages}{211--217}.
\bibitem[{Sharma et~al.(2024)Sharma, Gero, Payne, Gruber, Rus, Torralba and
  Andreas}]{sharma2024contextual}
\bibinfo{author}{Sharma, P.}, \bibinfo{author}{Gero, S.},
  \bibinfo{author}{Payne, R.}, \bibinfo{author}{Gruber, D.F.},
  \bibinfo{author}{Rus, D.}, \bibinfo{author}{Torralba, A.},
  \bibinfo{author}{Andreas, J.}, \bibinfo{year}{2024}.
\newblock \bibinfo{title}{Contextual and combinatorial structure in sperm whale
  vocalisations}.
\newblock \bibinfo{journal}{Nature Communications} \bibinfo{volume}{15},
  \bibinfo{pages}{3617}.
\bibitem[{Shklovskii and Sagan(1966)}]{shklovskii1966intelligent}
\bibinfo{author}{Shklovskii, I.S.}, \bibinfo{author}{Sagan, C.},
  \bibinfo{year}{1966}.
\newblock \bibinfo{title}{{Intelligent life in the Universe}}.
\newblock \bibinfo{publisher}{Holden-Day}.
\bibitem[{Smith(2000)}]{smith2000concept}
\bibinfo{author}{Smith, J.M.}, \bibinfo{year}{2000}.
\newblock \bibinfo{title}{The concept of information in biology}.
\newblock \bibinfo{journal}{Philosophy of science} \bibinfo{volume}{67},
  \bibinfo{pages}{177--194}.
\bibitem[{Soare(2009)}]{soare2009turing}
\bibinfo{author}{Soare, R.I.}, \bibinfo{year}{2009}.
\newblock \bibinfo{title}{Turing oracle machines, online computing, and three
  displacements in computability theory}.
\newblock \bibinfo{journal}{Annals of Pure and Applied Logic}
  \bibinfo{volume}{160}, \bibinfo{pages}{368--399}.
\bibitem[{Suddendorf and Corballis(2007)}]{suddendorf2007evolution}
\bibinfo{author}{Suddendorf, T.}, \bibinfo{author}{Corballis, M.C.},
  \bibinfo{year}{2007}.
\newblock \bibinfo{title}{The evolution of foresight: What is mental time
  travel, and is it unique to humans?}
\newblock \bibinfo{journal}{Behavioral and Brain Sciences}
  \bibinfo{volume}{30}, \bibinfo{pages}{299--313}.
\bibitem[{Svahn and Prokopenko(2023)}]{svahn2023ansatz}
\bibinfo{author}{Svahn, A.J.}, \bibinfo{author}{Prokopenko, M.},
  \bibinfo{year}{2023}.
\newblock \bibinfo{title}{{An Ansatz for computational undecidability in RNA
  automata}}.
\newblock \bibinfo{journal}{Artificial Life} \bibinfo{volume}{29},
  \bibinfo{pages}{261--288}.
\bibitem[{Szathm{\'a}ry(2015)}]{szathmary2015toward}
\bibinfo{author}{Szathm{\'a}ry, E.}, \bibinfo{year}{2015}.
\newblock \bibinfo{title}{Toward major evolutionary transitions theory 2.0}.
\newblock \bibinfo{journal}{Proceedings of the National Academy of Sciences}
  \bibinfo{volume}{112}, \bibinfo{pages}{10104--10111}.
\bibitem[{Szathm{\'a}ry and Maynard~Smith(1995)}]{szathmary1995major}
\bibinfo{author}{Szathm{\'a}ry, E.}, \bibinfo{author}{Maynard~Smith, J.},
  \bibinfo{year}{1995}.
\newblock \bibinfo{title}{The major evolutionary transitions}.
\newblock \bibinfo{journal}{Nature} \bibinfo{volume}{374},
  \bibinfo{pages}{227--232}.
\bibitem[{Szolnoki and Perc(2012)}]{szolnoki2012conditional}
\bibinfo{author}{Szolnoki, A.}, \bibinfo{author}{Perc, M.},
  \bibinfo{year}{2012}.
\newblock \bibinfo{title}{Conditional strategies and the evolution of
  cooperation in spatial public goods games}.
\newblock \bibinfo{journal}{Physical Review E} \bibinfo{volume}{85},
  \bibinfo{pages}{026104}.
\bibitem[{Szostak et~al.(2001)Szostak, Bartel and
  Luisi}]{szostak2001synthesizing}
\bibinfo{author}{Szostak, J.W.}, \bibinfo{author}{Bartel, D.P.},
  \bibinfo{author}{Luisi, P.L.}, \bibinfo{year}{2001}.
\newblock \bibinfo{title}{Synthesizing life}.
\newblock \bibinfo{journal}{Nature} \bibinfo{volume}{409},
  \bibinfo{pages}{387--390}.
\bibitem[{Tang et~al.(2020)Tang, Nguyen and Ha}]{tang2020neuroevolution}
\bibinfo{author}{Tang, Y.}, \bibinfo{author}{Nguyen, D.}, \bibinfo{author}{Ha,
  D.}, \bibinfo{year}{2020}.
\newblock \bibinfo{title}{Neuroevolution of self-interpretable agents}, in:
  \bibinfo{booktitle}{Proceedings of the 2020 Genetic and Evolutionary
  Computation Conference}, pp. \bibinfo{pages}{414--424}.
\bibitem[{Tomasello et~al.(2005)Tomasello, Carpenter, Call, Behne and
  Moll}]{tomasello2005understanding}
\bibinfo{author}{Tomasello, M.}, \bibinfo{author}{Carpenter, M.},
  \bibinfo{author}{Call, J.}, \bibinfo{author}{Behne, T.},
  \bibinfo{author}{Moll, H.}, \bibinfo{year}{2005}.
\newblock \bibinfo{title}{Understanding and sharing intentions: The origins of
  cultural cognition}.
\newblock \bibinfo{journal}{Behavioral and Brain Sciences}
  \bibinfo{volume}{28}, \bibinfo{pages}{675--691}.
\bibitem[{Trappes et~al.(2022)Trappes, Nematipour, Kaiser, Krohs, Van~Benthem,
  Ernst, Gadau, Korsten, Kurtz, Schielzeth et~al.}]{trappes2022individualized}
\bibinfo{author}{Trappes, R.}, \bibinfo{author}{Nematipour, B.},
  \bibinfo{author}{Kaiser, M.I.}, \bibinfo{author}{Krohs, U.},
  \bibinfo{author}{Van~Benthem, K.J.}, \bibinfo{author}{Ernst, U.R.},
  \bibinfo{author}{Gadau, J.}, \bibinfo{author}{Korsten, P.},
  \bibinfo{author}{Kurtz, J.}, \bibinfo{author}{Schielzeth, H.}, et~al.,
  \bibinfo{year}{2022}.
\newblock \bibinfo{title}{How individualized niches arise: defining mechanisms
  of niche construction, niche choice, and niche conformance}.
\newblock \bibinfo{journal}{BioScience} \bibinfo{volume}{72},
  \bibinfo{pages}{538--548}.
\bibitem[{Tria et~al.(2026)}]{tria2026language}
\bibinfo{author}{Tria, F.}, et~al., \bibinfo{year}{2026}.
\newblock \bibinfo{title}{Language games from a physics perspective}, in:
  \bibinfo{booktitle}{International Encyclopedia of Language and Linguistics}.
  \bibinfo{publisher}{Elsevier}, pp. \bibinfo{pages}{1--7}.
\bibitem[{Turing(1939)}]{turing1939systems}
\bibinfo{author}{Turing, A.M.}, \bibinfo{year}{1939}.
\newblock \bibinfo{title}{Systems of logic based on ordinals}.
\newblock \bibinfo{journal}{Proceedings of the London Mathematical Society,
  Series 2} \bibinfo{volume}{45}, \bibinfo{pages}{161--228}.
\bibitem[{Vansina(1985)}]{vansina1985oral}
\bibinfo{author}{Vansina, J.M.}, \bibinfo{year}{1985}.
\newblock \bibinfo{title}{Oral tradition as history}.
\newblock \bibinfo{publisher}{University of Wisconsin Press}.
\bibitem[{Vetsigian et~al.(2006)Vetsigian, Woese and
  Goldenfeld}]{vetsigian2006collective}
\bibinfo{author}{Vetsigian, K.}, \bibinfo{author}{Woese, C.},
  \bibinfo{author}{Goldenfeld, N.}, \bibinfo{year}{2006}.
\newblock \bibinfo{title}{Collective evolution and the genetic code}.
\newblock \bibinfo{journal}{Proceedings of the National Academy of Sciences}
  \bibinfo{volume}{103}, \bibinfo{pages}{10696--10701}.
\bibitem[{W{\"a}chtersh{\"a}user(1988)}]{wachtershauser1988before}
\bibinfo{author}{W{\"a}chtersh{\"a}user, G.}, \bibinfo{year}{1988}.
\newblock \bibinfo{title}{Before enzymes and templates: theory of surface
  metabolism}.
\newblock \bibinfo{journal}{Microbiological Reviews} \bibinfo{volume}{52},
  \bibinfo{pages}{452--484}.
\bibitem[{Wakano et~al.(2009)Wakano, Nowak and Hauert}]{wakano2009spatial}
\bibinfo{author}{Wakano, J.Y.}, \bibinfo{author}{Nowak, M.A.},
  \bibinfo{author}{Hauert, C.}, \bibinfo{year}{2009}.
\newblock \bibinfo{title}{Spatial dynamics of ecological public goods}.
\newblock \bibinfo{journal}{Proceedings of the National Academy of Sciences}
  \bibinfo{volume}{106}, \bibinfo{pages}{7910--7914}.
\bibitem[{Walker and Davies(2013)}]{walker2013algorithmic}
\bibinfo{author}{Walker, S.I.}, \bibinfo{author}{Davies, P.C.W.},
  \bibinfo{year}{2013}.
\newblock \bibinfo{title}{The algorithmic origins of life}.
\newblock \bibinfo{journal}{Journal of the Royal Society Interface}
  \bibinfo{volume}{10}, \bibinfo{pages}{20120869}.
\bibitem[{West et~al.(2015)West, Fisher, Gardner and Kiers}]{west2015major}
\bibinfo{author}{West, S.A.}, \bibinfo{author}{Fisher, R.M.},
  \bibinfo{author}{Gardner, A.}, \bibinfo{author}{Kiers, E.T.},
  \bibinfo{year}{2015}.
\newblock \bibinfo{title}{Major evolutionary transitions in individuality}.
\newblock \bibinfo{journal}{Proceedings of the National Academy of Sciences}
  \bibinfo{volume}{112}, \bibinfo{pages}{10112--10119}.
\bibitem[{Wiedmann and Lenzen(2018)}]{wiedmann2018environmental}
\bibinfo{author}{Wiedmann, T.}, \bibinfo{author}{Lenzen, M.},
  \bibinfo{year}{2018}.
\newblock \bibinfo{title}{Environmental and social footprints of international
  trade}.
\newblock \bibinfo{journal}{Nature Geoscience} \bibinfo{volume}{11},
  \bibinfo{pages}{314--321}.
\bibitem[{{Wikipedia contributors}(2024a)}]{wiki-Drawing_Hands}
\bibinfo{author}{{Wikipedia contributors}}, \bibinfo{year}{2024}a.
\newblock \bibinfo{title}{Drawing hands --- {W}ikipedia{,} the free
  encyclopedia}.
\newblock \URLprefix \url{https://en.wikipedia.org/wiki/Drawing_Hands}.
  \bibinfo{note}{[Online; accessed 21-June-2024]}.
\bibitem[{{Wikipedia contributors}(2024b)}]{wiki-Penrose_stairs}
\bibinfo{author}{{Wikipedia contributors}}, \bibinfo{year}{2024}b.
\newblock \bibinfo{title}{Penrose stairs --- {W}ikipedia{,} the free
  encyclopedia}.
\newblock \URLprefix \url{https://en.wikipedia.org/wiki/Penrose_stairs}.
  \bibinfo{note}{[Online; accessed 21-June-2024]}.
\bibitem[{{Wikipedia contributors}(2026)}]{wiki-ST}
\bibinfo{author}{{Wikipedia contributors}}, \bibinfo{year}{2026}.
\newblock \bibinfo{title}{Short-termism --- {W}ikipedia{,} the free
  encyclopedia}.
\newblock \URLprefix \url{https://en.wikipedia.org/wiki/Short-termism}.
  \bibinfo{note}{[Online; accessed 11-January-2026]}.
\bibitem[{Woese(1965)}]{woese1965evolution}
\bibinfo{author}{Woese, C.R.}, \bibinfo{year}{1965}.
\newblock \bibinfo{title}{On the evolution of the genetic code.}
\newblock \bibinfo{journal}{Proceedings of the National Academy of Sciences}
  \bibinfo{volume}{54}, \bibinfo{pages}{1546--1552}.
\bibitem[{Woese(1972)}]{woese1972emergence}
\bibinfo{author}{Woese, C.R.}, \bibinfo{year}{1972}.
\newblock \bibinfo{title}{The emergence of genetic organization}, in:
  \bibinfo{editor}{Ponnamperuma, C.} (Ed.), \bibinfo{booktitle}{Exobiology}.
  \bibinfo{publisher}{North-Holland Publishing Co., Amsterdam, The
  Netherlands}, pp. \bibinfo{pages}{301--341}.
\bibitem[{Woese(2000)}]{woese2000interpreting}
\bibinfo{author}{Woese, C.R.}, \bibinfo{year}{2000}.
\newblock \bibinfo{title}{Interpreting the universal phylogenetic tree}.
\newblock \bibinfo{journal}{Proceedings of the National Academy of Sciences}
  \bibinfo{volume}{97}, \bibinfo{pages}{8392--8396}.
\bibitem[{Woese(2004)}]{woese2004new}
\bibinfo{author}{Woese, C.R.}, \bibinfo{year}{2004}.
\newblock \bibinfo{title}{A new biology for a new century}.
\newblock \bibinfo{journal}{Microbiology and molecular biology reviews}
  \bibinfo{volume}{68}, \bibinfo{pages}{173--186}.
\bibitem[{Yanofsky(2003)}]{yanofsky2003universal}
\bibinfo{author}{Yanofsky, N.S.}, \bibinfo{year}{2003}.
\newblock \bibinfo{title}{A universal approach to self-referential paradoxes,
  incompleteness and fixed points}.
\newblock \bibinfo{journal}{Bulletin of Symbolic Logic} \bibinfo{volume}{9},
  \bibinfo{pages}{362--386}.
\bibitem[{Yockey(1992)}]{yockey1992information}
\bibinfo{author}{Yockey, H.P.}, \bibinfo{year}{1992}.
\newblock \bibinfo{title}{Information theory and molecular biology}.
\newblock \bibinfo{publisher}{Cambridge University Press}.
\bibitem[{Zahavi(1975)}]{zahavi1975mate}
\bibinfo{author}{Zahavi, A.}, \bibinfo{year}{1975}.
\newblock \bibinfo{title}{Mate selection—a selection for a handicap}.
\newblock \bibinfo{journal}{Journal of Theoretical Biology}
  \bibinfo{volume}{53}, \bibinfo{pages}{205--214}.

\end{thebibliography}

\end{document}